\documentclass[structabstract]{aa}

\usepackage{amsmath}          
\usepackage{dcolumn}          
\usepackage{graphicx}         
\usepackage{txfonts}          
\usepackage{enumitem}
\usepackage{mathrsfs}         
\usepackage{rotating}         
\usepackage{subfigure}        
\usepackage{lscape}           
\usepackage{afterpage}        
\usepackage{xspace}           
\usepackage{natbib}           
\usepackage[version=4]{mhchem}
\usepackage{xymtex}

\providecommand*{\arcsec}{\ensuremath{^{\prime\!\prime}}\xspace}

\providecommand*{\hms}[3]{\ensuremath{{#1}^\mrm{h}{#2}^\mrm{m}{#3}^\mrm{s}}\xspace}
\providecommand*{\dms}[3]{\ensuremath{#1^\circ #2^\prime #3^{\prime\prime}}\xspace}

\newcommand{\mcl}[3]{\multicolumn{#1}{#2}{#3}}
\newcommand{\mrm}[1]{\ensuremath{\mathrm{#1}}}
\newcommand{\bsy}[1]{\ensuremath{\boldsymbol{#1}}}

\newchemenvironment{chemalignat}{alignat}
\newchemenvironment{chemgather}{gather}
\newcolumntype{.}{D{.}{.}{-1}}

\begin{document}

\title{Propargylimine in the laboratory and in space: \\
       millimetre-wave spectroscopy and first detection in the ISM}

\author{L.~Bizzocchi\inst{\ref{MPE}} \and
        D.~Prudenzano\inst{\ref{MPE}} \and 
        V.~M.~Rivilla\inst{\ref{Arcetri}} \and
        A.~Pietropolli-Charmet\inst{\ref{CaFo}} \and
        B.~M.~Giuliano\inst{\ref{MPE}} \and
        P.~Caselli\inst{\ref{MPE}} \and
        J.~Mart\'in-Pintado\inst{\ref{CAB}} \and
        I.~Jim\'enez-Serra\inst{\ref{CAB}} \and
        S.~Mart\'in\inst{\ref{ESO-C},\ref{ALMA-J}} \and
        M.~A.~Requena-Torres\inst{\ref{Mary}, \ref{Tow}} \and
        F.~Rico-Villas\inst{\ref{CAB}} \and
        S.~Zeng\inst{\ref{RIKEN}} \and
        J.-C.~Guillemin\inst{\ref{Ren}}}

\institute{Center for Astrochemical Studies, 
           Max-Planck-Institut f\"ur extraterrestrische Physik,
           Gie\ss enbachstra\ss e 1, 85748 Garching (Germany)
           \email{[prudenzano,bizzocchi,giuliano,caselli]@mpe.mpg.de}
           \label{MPE}
           \and
           INAF-Osservatorio Astrofisico di Arcetri,
           Largo Enrico Fermi 5, 50125 Firenze (Italy)
           \email{rivilla@arcetri.astro.it}
           \label{Arcetri}
           \and
           Dipartimento di Scienze Molecolari e Nanosistemi, 
           Universit\`a Ca'~Foscari Venezia, via Torino~155, 30172 Mestre (Italy)
           \email{jacpnike@unive.it}
           \label{CaFo}
           \and
           Centro de Astrobiolog\'ia (CSIC--INTA), 
           Ctra. de Torrej\'on a Ajalvir, km~4, Torrej\'on de Ardoz, 28850 Madrid (Spain)
           \label{CAB}
           \and
           European Southern Observatory (ESO),
           Alonso de C\'ordoba 3107, Vitacura, 763-0355 Santiago (Chile)
           \label{ESO-C}
           \and
           Joint ALMA Observatory (ESO),
           Alonso de C\'ordoba 3107, Vitacura, 763-0355 Santiago (Chile)
           \label{ALMA-J}
           \and
           University of Maryland,
           College Park, ND 20742-2421 (USA)
           \label{Mary}
           \and
           Department of Physics, Astronomy and Geosciences, Towson University, 
           Towson, MD 21252, USA
           \label{Tow}
           \and
           Star and Planet Formation Laboratory, Cluster for Pioneering Research, RIKEN,
           2-1 Hirosawa, Wako, Saitama, 351-0198, Japan
           \email{shaoshan.zeng@riken.jp}
           \label{RIKEN}
           \and
           Univ. Rennes, Ecole Nationale Sup\'erieure de Chimie de Rennes, 
           CNRS, ISCR -- UMR6226, F-35000 Rennes (France)
           \email{jean-claude.guillemin@ensc-rennes.fr}
           \label{Ren}
           }

\titlerunning{Propargylimine in laboratory and in space}
\authorrunning{Prudenzano et al.}
%

\abstract
%
{
Small imines containing up to three carbon atoms are present in the interstellar medium.
As alkynyl compounds are abundant in this medium, propargylimine (2-propyn-1-imine,
HC$\equiv$C$-$CH$=$NH) thus represents a promising candidate for a new interstellar detection.
}
%
{
The goal of the present work is to perform a comprehensive laboratory investigation of the 
rotational spectrum of propargylimine in its ground vibrational state in order to obtain 
a highly precise set of rest-frequencies and to search it in space.
}
%
{
The rotational spectra of $E$ and $Z$ geometrical isomers of propargylimine have been 
recorded in laboratory in the 83--500\,GHz frequency interval.
The measurements have been performed using a source-modulation millimetre-wave spectrometer
equipped with a pyrolysis system for the production of unstable species.
High-level ab initio calculations were performed to assist the analysis and to obtain 
reliable estimates for an extended set of spectroscopic quantities.
We have searched for propargylimine at~3 and 2\,mm in the spectral survey of the quiescent 
giant molecular cloud G+0.693-0.027 located in the ``Central Molecular Zone'', close to the 
Galactic Centre.
}
%
{
About 1000 rotational transitions have been recorded for the $E$- and $Z$-propargylimine, 
in the laboratory.
These new data have enabled the determination of a very accurate set of spectroscopic
parameters including rotational, quartic and sextic centrifugal distortion constants.
The improved spectral data allow us to perform a successful search for this new imine
in the G+0.693-0.027 molecular cloud.
Eighteen lines of $Z$-propargylimine have been detected at level $>2.5$\,$\sigma$, resulting
in a column density estimate of $N = (0.24\pm 0.02)\times 10^{14}$\,cm$^{-2}$\@.
An upper limit was retrieved for the higher-energy $E$ isomer, which was not detected in the 
data.
The fractional abundance (w.r.t.~\ce{H2}) derived for $Z$-propargylimine is $1.8\times 10^{-10}$.
We discuss the possible formation routes by comparing the derived abundance with those 
measured in the source for possible chemical precursors.
}
%
{}


\maketitle
%

\section{Introduction} \label{sec:intro}
\indent\indent
Among the over 200~molecules detected in the Interstellar Medium (ISM) and circumstellar 
shells, approximately~70 have 6~or more atoms and contain carbon. 
In an astronomical context, these compounds are called Complex Organic Molecules 
\citep[COMs,][]{Herbst-ARAA09-COMs}, and have received an increasing attention in the last 
decade in an effort of unveiling how chemical complexity builds up, 
from simple species to the much larger molecular structures required to establish biochemical
processes on planets where appropriate conditions are met.
Nitrogen-bearing COMs are particularly interesting as they can be regarded as an 
intermediate step towards the formation of biologically important species, such as 
nucleobases and amino-acids.
As protein constituents, amino acids are prime targets for astrobiological studies, and 
their extra-terrestrial formation has been a highly debated topic.
Numerous compounds of this class have been found in carbonaceous chondrites 
\citep[see e.g.][and references therein]{Cobb-ApJ14-AAs}, where they are thought to form 
by aqueous alteration (see \citealt{Burton-CSR12-AAs} for a review on meteoritic 
prebiotic compounds).
Glycine, the simplest amino acid, has not been detected in the ISM to date
\citep{Snyder-ApJ05-Gly}, however it was found in the coma of the 67P/Churyumov--Gerasimenko 
comet through \textit{in situ} mass spectrometry \citep{Altwegg-AS16-P67}.
The presence of glycine in the volatile cometary material thus strongly suggests the 
existence of a process capable to generate amino acids in cold environments and likely in 
absence of liquid water.

\begin{figure*}[h!]
 \begin{minipage}{0.5\textwidth}
  \centering
  \includegraphics[angle=0,width=0.95\textwidth]{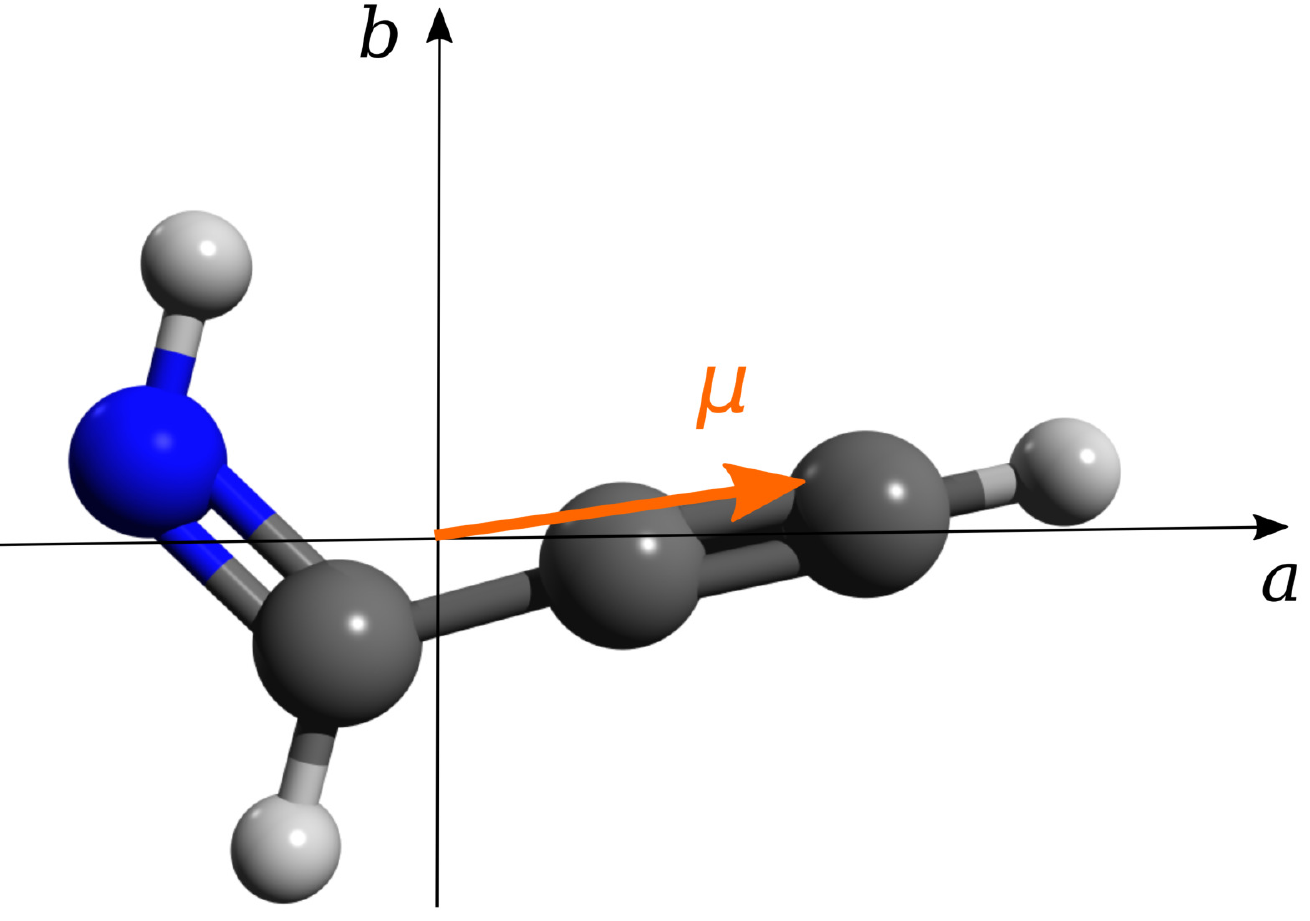} \\[1ex]
  {\Large $Z$-PGIM}
 \end{minipage}
 \begin{minipage}{0.5\textwidth}
  \centering
  \includegraphics[angle=0,width=0.95\textwidth]{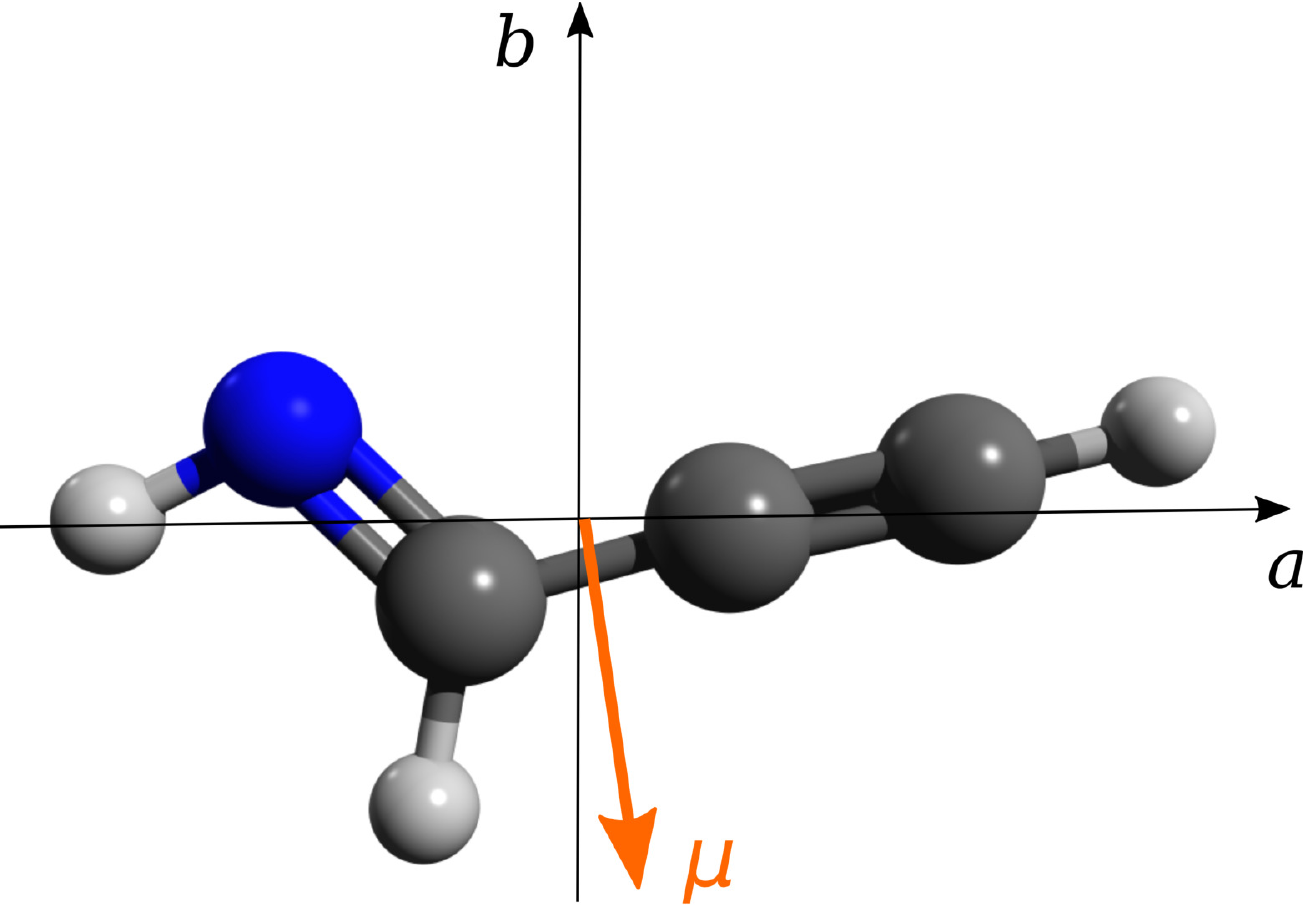} \\[1ex]
  {\Large $E$-PGIM}
 \end{minipage}
 \caption{Molecular structure and principal inertial axes of the $Z$ and $E$ isomers
          of PGIM.
          The orange arrow indicates the direction of the electric dipole moment 
          $\boldsymbol{\mu}$ and points towards the displacement of the notional 
          negative charge.
          The modulus is $\mu = 2.15$\,D for $Z$-PGIM and $\mu = 1.95$\,D for $E$-PGIM.
          \label{fig:molec}}
\end{figure*}

Many theoretical and laboratory studies have been devoted to the investigation of the 
chemical routes which may lead to amino acids in diverse extra-terrestrial environments
\citep[see e.g.][and references therein]{Woon-ApJ02-Aacids,Koch-JPC08-AmAcCN,Aponte-ACS17-Pathw}.
The most promising pathways in interstellar ice analogues involve, as the last step, 
the hydration of an amino-nitrile compound (\ce{H2N-CHR-CN}).
In astrophysical-like conditions, these precursors may be generated photochemically, i.e. 
through addition of ammonia to the corresponding nitrile \citep{Danger-AA11-PhotChem}, 
or via the Strecker mechanism which starts from the condensation of ammonia with an aldehyde 
\citep[\ce{R-CHO}, see ][and references therein]{Danger-AA11-StreckSyn}.
This latter process involves a species containing the iminic moiety (\ce{RC=NH}) 
as reactive intermediate \citep{Aponte-ACS17-Pathw}.

Imines are a class of molecules that are well represented in the ISM with six members 
detected to date. 
Four are simple chains: 
methanimine \citep[\ce{CH2NH},][]{Godfrey-ApL73-MetIm,Dick-ApJ97-MetIm}, 
ethanimine \citep[\ce{CH3CHNH},][]{Loomis-ApJ13-EthIm},
ketenimine \citep[\ce{CH2CNH},][]{Lovas-ApJ06-KetIm}, and 
3-imino-1,2-propadienylidene \citep[\ce{CCCNH},][]{Kawa-ApJ92-C3NH};
but there are also the substituted 
$C$-cyanomethanimine \citep[\ce{NCCHNH},][]{Zaleski-ApJ13-CNMtIm,Rivilla-MNRAS19-CNMtIm}, 
and the cumulated ``diimine'' carbodiimide \citep[\ce{HNCNH},][]{McGuire-ApJ12-CdIm}.
Hypotheses on their formation in astrophysical environments point mainl 	  y to a chemical 
route from simple nitriles via tautomerisation \citep{Lovas-ApJ06-KetIm} 
or by partial hydrogenation on dust grain surface 
\citep{Theule-AA11-hydHCN,Krim-MNRAS19-MKetIm}.

According to this scheme, it results that, among the 3C-atom bearing imines, a new, promising 
candidate for the detection in the ISM would be propargylimine 
(2-propyn-1-imine, \ce{HC#C-CH=NH}, hereafter PGIM).
This species is a structural isomer of the well-known astrophysical molecule acrylonitrile
\citep{Gardner-ApJ75-AcrCN} and can be chemically related --- through 2H addition --- 
to cyanoacetylene (\ce{HC3N}), a nitrile species which is ubiquitous in the ISM 
\citep[e.g.][and references therein]{Bizz-ApJS17-HC3N}. 
Despite this, PGIM has not attracted a great deal of interest from laboratory 
spectroscopists and only a few, rather outdated works, are present in the literature.
Its rotational spectrum was first observed in \citeyear{Kroto-JCS84-PGIM} by 
\citeauthor{Kroto-JCS84-PGIM} in the centimetre-wave (cm-wave) spectral region. 
Shortly after, the study was extended by \citet{Sugie-JMS85-PGIM} and by 
\citet{McNaugh-JMSt88-PGIM}, who also recorded a few rotational transitions for several 
isotopic variants.
In the mid `80s, the observation of its low-resolution infrared (IR) spectrum was also 
reported \citep{Hamada-JMS84-PGIM,Osman-JCSpt87-PGIM}.
These few laboratory studies do not provide an exhaustive spectroscopic knowledge.
In particular, the coverage of the rotational spectrum is sparse and limited to the 
cm-wave regime, thus the reliability of the rest-frequency computed at millimetre (mm) 
wavelengths is not suitable for the purpose of an effective astronomical search.
With this in mind, we have undertaken a new extensive laboratory investigation, 
recording the mm spectrum of PGIM in its vibrational ground state.
The newly obtained laboratory data have been then used as a guidance to search for PGIM
towards the quiescent molecular clouds G+0.693-0.027, located in the ``Central Molecular Zone'' 
in the inner $\sim$500\,pc of our Galaxy, where several species directly related with 
prebiotic chemistry have been recently detected
\citep{ReqTorr-AA06-HCC,ReqTorr-ApJ08-OCOM,Zeng-MNRAS18-COMs,JimSerra-inpress-Urea}.

The structure of the paper is the following.
In Sect.~\ref{sec:exp} we describe the experimental procedure and in Sect.~\ref{sec:mol-prop} 
we provide a short account of the theoretical calculations performed to support the data 
analysis. 
Sect.~\ref{sec:anal} we descride the spectral analysis and discuss the results.
In Sect.~\ref{sec:detect} we describe the observations performed to search for 
PGIM in the ISM and illustrate the analysis of the data which leads to
its positive identification.
Finally, we draw our conclusions in Sect.~\ref{sec:conc}.

\section{Experiments} \label{sec:exp}
\indent\indent
The rotational spectrum of PGIM has been recorded using the CASAC 
(Center for Astrochemical Studies Absorption Cell) spectrometer at the 
Max-Planck-Institut f\"ur extraterrestrische Physik in Garching.
Full details on the experimental set-up have been already provided 
\citep{Bizz-AA17-HOCO+}; here, we report only a few key details which apply to the 
present investigation.
The instrument employs an active multiplier chain (Virginia Diodes) as a
source of mm radiation in the 82--125\,GHz band.
This primary stage is driven by a cm-wave synthesizer which operates in the 18-28\,GHz 
frequency range.
Accurate frequency and phase stabilisation is achieved by locking the parent cm synthesizer 
to a Rb atomic clock.
By the use of further multiplier stage in cascade, the frequency coverage can be extended 
towards the sub-mm regime with available power of a 2--20 $\mu$W up to 
1\,THz\@.
A closed-cycle He-cooled InSb hot electron bolometer operating at 4\,K (QMC) is used as 
a detector.
The spectral measurements have been performed using the frequency modulation (FM) technique:
the carrier signal from the cm-wave synthesizer is sine-wave modulated at 50\,kHz and the 
detector output is demodulated at twice this frequency ($2f$) by a lock-in amplifier.
The second derivative of the actual absorption profile is thus recorded by the 
computer-controlled acquisition system.

\begin{figure}[t]
 \centering
 \includegraphics[angle=0,width=0.43\textwidth]{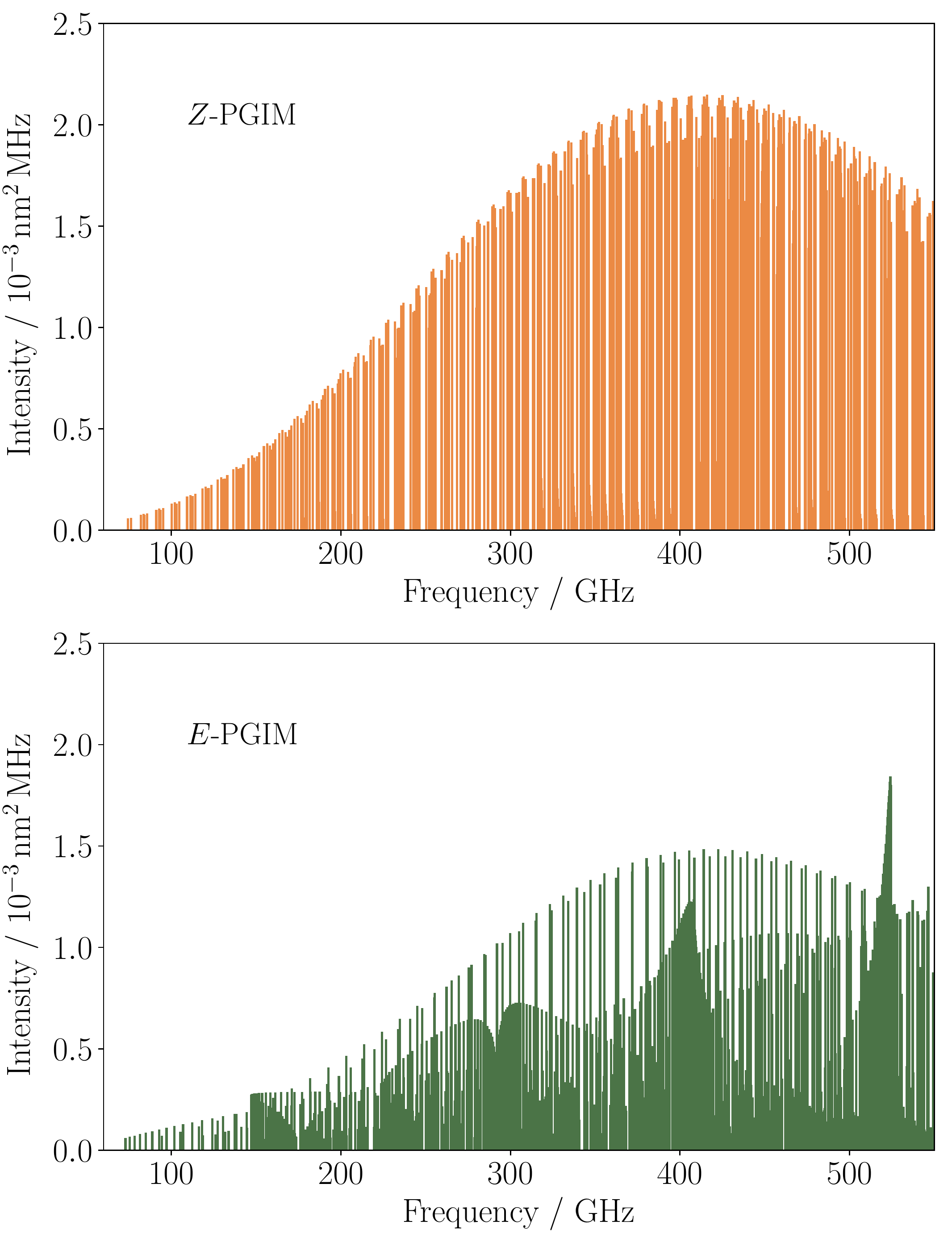}
 \caption{Stick spectra of the $Z$ (\textit{upper panel}) and $E$ (\textit{lower panel})
             isomers of PGIM computed at 300\,K\@.
             Frequency positions and intensities have been computed using the spectroscopic
             parameters reported in Table~\ref{tab:specpar} and the ab initio values of the 
             dipole moments (see Appendix~\ref{sec:theor-details}).
        \label{fig:stick-spec}}
\end{figure}

The absorption cell is a plain Pyrex tube (3\,m long and 5\,cm in diameter) which, at one 
end is connected to a side arm hosting a pyrolysis production system.
This consists of a quartz tube (1\,cm in diameter, 60\,cm long) inserted in a tubular oven
(Carbolite) which heats up the inner 40\,cm long part.

PGIM was produced as in \citet{Sugie-JMS85-PGIM}, i.e., by pyrolysing dipropargylamine 
(\ce{(HC\bond{3}CCH2)2NH}) vapours and flowing the gaseous reaction products through the 
absorption cell kept under continuous pumping. 
In our setup, the strongest absorption signals of the target molecule were obtained 
by setting the oven temperature at 950$^\circ$C\@.
Typical pressure were 150-200\,mTorr (20--26\,Pa) at the quartz tube inlet. 
This corresponds to ca.~4\,mTorr (0.5\,Pa) in the absorption cell which was kept at room temperature.
Multiple side-products are formed, as reported by \citet{McNaugh-JMSt88-PGIM}, which
occasionally generated strong spectral features. 
This, however did not hamper the recording of the PGIM spectrum, thus sample purification
via selective trapping or condensation/re-vaporisation were not attempted in the present 
investigation.

\section{Molecular properties} \label{sec:mol-prop}
\indent\indent
PGIM can be described as a molecule joining two basic subunits: the ethynyl group 
\ce{HC\bond{3}C\bond{1}} and the iminic moiety \ce{-CH=NH}.
The presence of two conjugated multiple bonds forces the molecule to the planar 
configuration with all the seven atoms lying on the plane defined by the $a$ and $b$ 
principal axes.
Owing to different relative position of the \ce{HCC} group and of the iminic H with 
respect to the \ce{C=N} double bond, two structural isomers exist: $Z$ and $E$.
Their structures are depicted in Fig.~\ref{fig:molec}.

With the aim of fully characterising the molecular properties of the PGIM we have conducted 
extensive theoretical calculations at coupled-cluster (CC) level of theory, with single 
and double excitations augmented by a perturbative treatment for estimating the effects of 
triple excitations \citep[i.e. CCSD(T),][]{Ragh-CPL89-CCSDT}.
Appropriate extrapolation procedures to the complete basis set (CBS) limit 
\citep{Heckert-MP05-CCSDTQ,Heckert-JCP06-CCextp} were then employed to estimate the 
equilibrium structures of the two conformers and their energies.
Employing a similar composite approach 
\citep[see, for example][]{Barone-ACR15-QC,PietroPC-JPC17-BrFEt,CDE-AA18-PGAM} the 
best-estimate values of the quadratic force fields were derived. 
Cubic and semidiagonal quartic  force constants, computed at different levels of theory, 
were employed for the vibrational corrections to the equilibrium rotational constants, 
for the anharmonic corrections to the harmonic frequencies, and for determining the sextic 
centrifugal distortion constants. 
Nuclear quadrupole coupling constants for the nitrogen atoms were computed following the 
same procedure described previously \citep{Cazzoli-JPC11-CHBrF2,PietroPC-JPC16-ClFEt}. 
Additional details on the calculations (levels of theory, basis sets, and methodology) 
are reported in the Appendix~\ref{sec:theor-details}.

From our calculations the energy difference between the $Z$ (more stable) and $E$ isomers 
is 0.8510\,kcal\,mol$^{-1}$ ($E/k = 428.2$\,K)\@.
Both isomers are prolate-type slightly asymmetric tops ($\kappa\sim 0.98$). 
While the modulus of the dipole moment is similar ($|\bsy{\mu}|\sim 2$\,D), 
the corresponding components in the principal axes are very different: 
$\mu_a = 2.14$\,D and $\mu_b = 0.17$\,D for the $Z$ isomer; 
$\mu_a = 0.26$\,D and $\mu_b = 1.93$\,D for the $E$ isomer.
The dipole moment vectors are also shown in Fig.~\ref{fig:molec}.

\begin{table*}[h!]
 \caption[]{Summary of the transitions with resolved hyperfine structure recorded for
            both PGIM isomers.
            \label{tab:hfs-summary}}
 \footnotesize
 \begin{center}
 \begin{tabular}{l ccc c l ccc}
  \hline  \\[-1ex]
   \mcl{4}{c}{$Z$-PGIM}  &  & \mcl{4}{c}{$E$-PGIM} \\[0.5ex]
   \cline{1-4}\cline{6-9} \\[-1ex]
   type$^a$ & \mcl{1}{c}{no. of lines} & no. of comp.$^b$ & $K_a'$ & & 
   type$^a$ & \mcl{1}{c}{no. of lines} & no. of comp.$^b$ & $K_a'$   \\[1ex]
  \hline  \\[-1ex]
  $^aR_{0,+1}$   &  29  &   76  &  0,1,2,3  & & $^bP_{+1,-1}$  &  10  &  18  &  1     \\[0.5ex]
  $^bR_{+1,+1}$  &   4  &   6   &  0        & & $^bR_{0,+1}$   &  73  & 140  &  0,1,2 \\[0.5ex]
  $^bR_{-1,+1}$  &   3  &   6   &  1        & & $^bQ_{+1,-1}$  &  50  &  99  &  0,1,2 \\[0.5ex]
  $^bQ_{+1,-1}$  &   6  &  12   &  0        & & \\[0.5ex]
  \hline\hline \\[-2ex]
 \end{tabular}
 \end{center}
 $^a$ See footnote at page~\pageref{fn:tlab}. \\
 $^b$ Resolved hyperfine components.
\end{table*}

\section{Spectral analysis} \label{sec:anal}
\indent\indent
The rotational spectrum of PGIM has been recorded in selected frequency intervals from~83 
to 500\,GHz.
In the mm region, $Z$-PGIM presents a typical $a$-type spectrum with groups of $R$-branch 
$\Delta K_a = 0$ transitions regularly separated by $\approx B+C$, (Fig.~\ref{fig:stick-spec}, 
\textit{upper panel}), while the $E$ isomer exhibits a much more complex spectrum consisting 
of several $\Delta K_a \pm 1$ ladders overlapped with each other and with some prominent 
$Q$-branch band-heads spaced by $\approx 2A-B-C$ (Fig.~\ref{fig:stick-spec}, 
\textit{lower panel}).

From the ab initio computed energy difference, a relative $[Z]/[E]$ isomer abundance 
of~4.2 can be estimated at 300\,K\@.
From the relative intensity comparison between a pair of nearby $Z$- and $E$-PGIM lines 
recorded under the same experimental conditions (source power, sample pressure, and 
modulation depth) we obtained $[Z]/[E] = 4.7\pm 0.7$, in good agreement with the 
theoretical computation.
This indicates that, although PGIM is generated in a high-temperature environment 
($\sim 950$\,$^\circ$C), there is a quick thermalisation between the two isomers, 
and the population relaxes to the 300\,K value right after the gas enters in contact 
with the cell walls.

\begin{figure}[t]
 \centering
 \includegraphics[angle=0,width=0.42\textwidth]{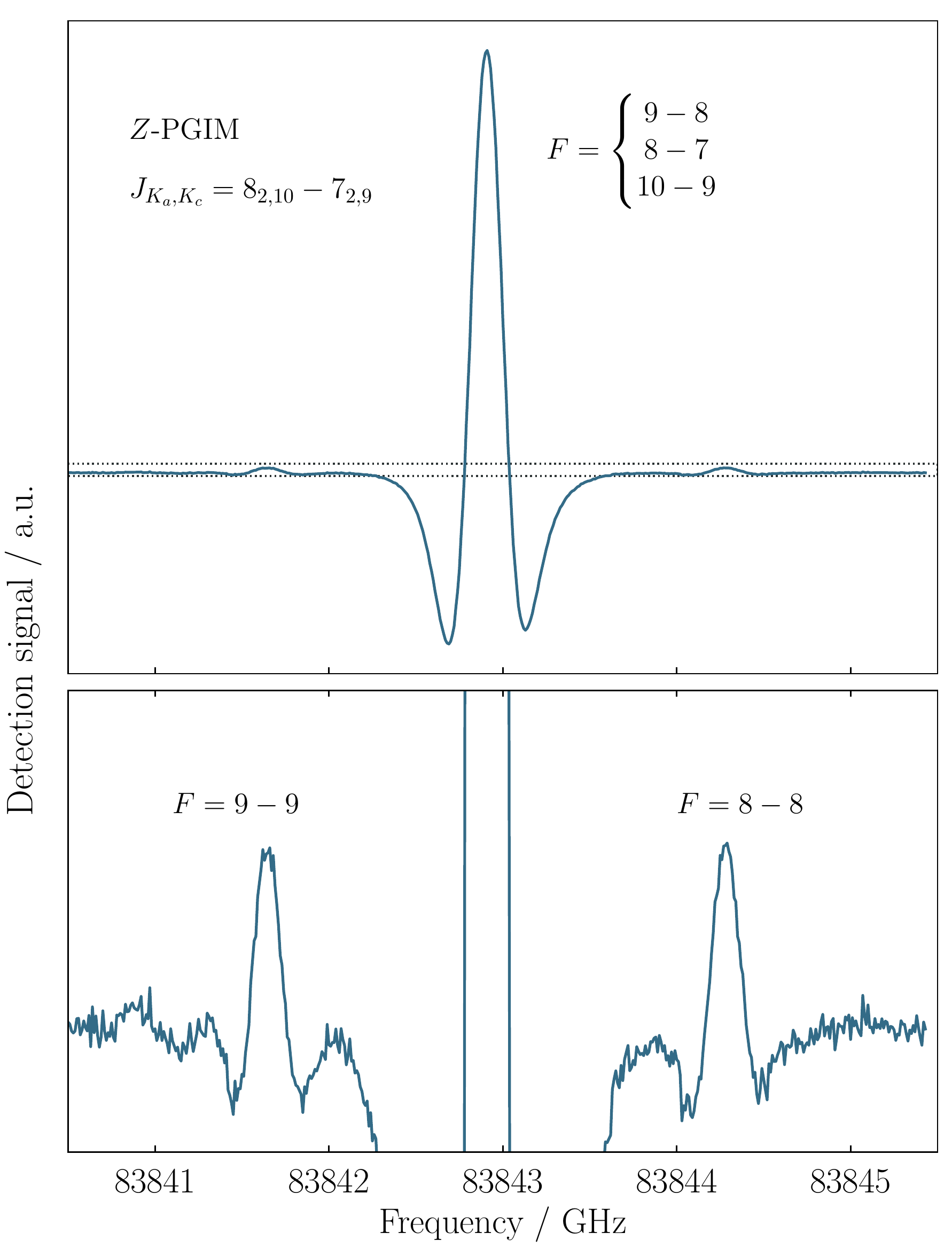}
 \caption{Recording of the $J_{K_a,K_c} = 8_{2,10}-7_{2,9}$ of $Z$-PGIM showing the
          two weak $\Delta F = 0$ hyperfine components symmetrically separated from the
          central blended  $\Delta F = +1$ triplet by 1.3 MHz (\textit{upper panel}). 
          Total integration time 47\,s with time constant $RC = 3$\,ms.
          The area enclosed in the dashed box is plotted with expanded $y$-axis in the 
          \textit{lower panel}.
          \label{fig:DeltaF0}}
\end{figure}

\begin{figure*}[t]
 \centering
 \includegraphics[angle=0,width=1.0\textwidth]{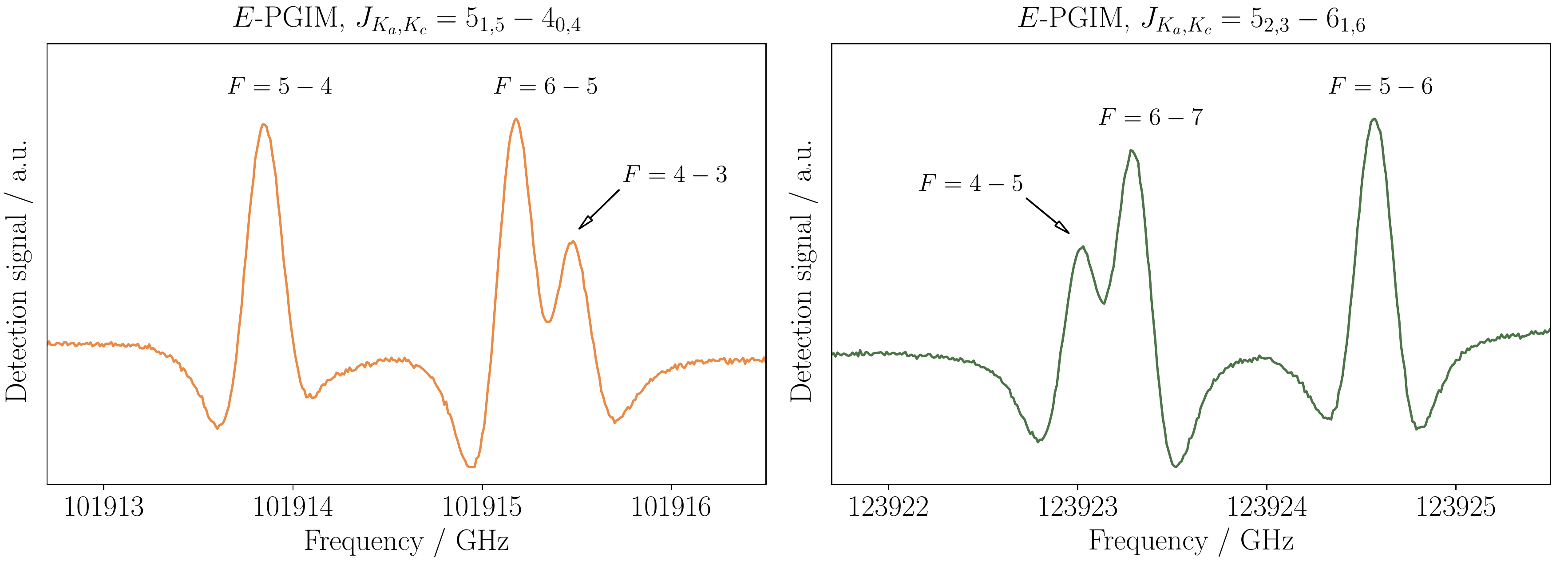}
 \caption{Recordings of two $b$-type transitions for $E$-PGIM showing the
          typical hyperfine structure produced by the quadrupole coupling of the $^{14}$N
          nucleus.
          \textit{Left panel}: $J_{K_a,K_c} = 5_{1,5} - 4_{0,4}$; integration time 180\,s.
          \textit{Right panel}: $J_{K_a,K_c} = 5_{2,3} - 6_{1,6}$; integration time 165\,s.
          The adopted scan rate is 0.2\,MHz\,s$^{-1}$ with time constant $RC = 3$\,ms\@.
          \label{fig:DeltaF1}}
\end{figure*}

Due to the presence of nitrogen, hyperfine coupling is generated between the molecular 
electric field gradient averaged over the end-over-end rotation, and the quadrupole 
moment of the $^{14}$N nucleus having spin $I=1$.
Thus, each rotational level with principal quantum number $J>0$ is split into three 
hyperfine sublevels labelled with the total angular quantum number $F$, where 
$F=J-1,J,J+1$\@.
As a consequence, the transitions are split into several components according to the 
$\Delta F=0,\pm 1$ selection rules, with the strongest features being those for which 
$\Delta F=\Delta J$\@.
However, in the frequency interval covered by the present investigation, the $J$ quantum 
number reaches a value as high as 54, thus most of the $^aR_{0,+1}$ transitions\footnote%
{    
    The symbol $^xM_{\delta K_a,\delta K_c}$ is used to label in a compact form the transition
    type for an asymmetric rotor: $x$ indicates the dipole moment component involved, 
    $M = P,Q,R$ is the symbol for the transitions with $\Delta J = -1,0,+1$, respectively, 
    and $\delta K_a$ and $\delta K_c$ refer to the (signed) change of the $K_a$ and $K_c$ 
    pseudo-angular quantum numbers \citep{Gordy-1984}\label{fn:tlab}.
}
(which dominates the $Z$ isomer spectrum) have their hyperfine pattern collapsed into 
a single feature. 
Nevertheless, for a few low-$J$ lines it was possible to detect the very weak $\Delta F=0$ 
components, which form a widely separated doublet approximately centred at the frequency of 
the corresponding unsplit transition.
An example of such hyperfine patterns is given in Fig.~\ref{fig:DeltaF0}.
The situation is different for the $b$-type lines, typical of the $E$ isomer spectrum. 
Due to the change in the $K_a$ pseudo quantum number involved in these transitions, less 
tight hyperfine patterns are produced and triplets/doublets of lines have been generally 
recorded, as shown in Fig.~\ref{fig:DeltaF1}. 
A summary of the transitions of both isomers for which the hyperfine structure have been 
resolved is presented in Table~\ref{tab:hfs-summary}.

Whenever possible, each measured feature was assigned to a single quadrupole component.
Close patterns were observed as a single line and, in this case, the intensity-averaged 
calculated frequency (typically of 2 or 3 components) was compared with the experimental 
datum in the least-squares fit.
Loose blends of unresolved components have also been observed.
These lines appeared as broad and distorted features and were not used in the analysis.
This careful selection procedure made it possible to achieve the same measurement 
precision for both the singly-assigned and for the intensity-averaged hyperfine entries.
A sizeable number of recorded transitions did not show any hint of hyperfine splitting, 
namely 381~for the $Z$ isomer and 288~for the $E$ isomer. 
For these data, the contributions due to the nuclear quadrupole coupling were neglected,
and the measured frequencies were assigned to the corresponding pure rotational 
transitions.
For each isomer, lines with or without resolved hyperfine structure have been given the 
same assumed uncertainty and have been analysed together in a global least-squares 
fashion.

In total, our data sets comprise 531~lines for $Z$-PGIM and 545~lines for 
$E$-PGIM. 
They also included 50~and, respectively, 47~cm-wave transitions taken from the 
literature \citep{Sugie-JMS85-PGIM,McNaugh-JMSt88-PGIM}.
Different statistical weights $w = 1/\sigma^2$ were given to the data of the various 
subsets to take into account the different measurement precision ($\sigma$).
For the lines measured by \citet{Sugie-JMS85-PGIM} we retained the same weighting 
scheme of the original paper, while 50\,kHz was given to the few transitions of
the $Z$ isomer reported by \citet{McNaugh-JMSt88-PGIM}.
The average uncertainty of the frequencies measured in the present work is estimated
to be 15\,kHz\@.
The complete data list is provided as electronic supplementary information.

The hyperfine energies were computed adopting the standard vector coupling scheme between
the rotational angular momentum $J$ and the nitrogen spin $I_\mrm{N}$:
\begin{equation}
\mathbf{J} + \mathbf{I}_\mrm{N} = \mathbf{F} \,.
\end{equation}
The total Hamiltonian is thus expressed as the sum of a purely rotational part and a
hyperfine contribution:
\begin{equation}
\hat{H} = \hat{H}_\text{rot} + \hat{H}_\text{HFS} \,
.\end{equation}
The pure rotational levels are labelled with the quantum numbers $J_{K_a,K_c}$, the total 
angular momentum quantum number $F$ must be added when the hyperfine sublevels are considered.
The rotational Hamiltonian $\hat{H}_\text{rot}$ is the $S$-reduced Watson-type Hamiltonian
in its I$^r$ representation (\citealt{Watson-1977}) and includes centrifugal distortion 
corrections up to the octic terms.
The hyperfine-structure Hamiltonian $\hat{H}_\text{HFS}$ is expressed by the traceless 
tensor $\boldsymbol{\chi}$ which has  $\chi_{aa}$ and $\chi_{bb}-\chi_{cc}$ as determinable 
coefficients.
This term has been considered only for the analysis of the resolved hyperfine components.
The weak spin--rotation couplings of both $^{14}$N and H nuclei do not produce any 
detectable effect in the recorded spectra and were thus neglected.
All spectral computations were performed with the CALPGM suite of programs 
\citep{Pick-JMS91-calpgm}.

\begin{table*}[t]
 \caption[]{Experimental and theoretical spectroscopic parameters of PGIM isomers. 
            Numbers in parentheses are 1$\sigma$ statistical uncertainties in the units 
            of the last quoted digit.
            \label{tab:specpar}}
 \footnotesize
 \begin{center}
 \begin{tabular}{ll D{.}{.}{10}D{.}{.}{4} c D{.}{.}{10}D{.}{.}{4}}
  \hline  \\[-1ex]
            &       &  \mcl{2}{c}{$Z$-PGIM} & &  \mcl{2}{c}{$E$-PGIM} \\[1ex]
  \cline{3-4} \cline{6-7} \\[-1ex]
  \mcl{2}{l}{Parameter} & \mcl{1}{c}{exp.} & \mcl{1}{c}{ab initio$^a$} & & \mcl{1}{c}{exp.} & \mcl{1}{c}{ab initio$^a$} \\[1ex]
  \hline  \\[-1ex]
  $A$       & / MHz &  54640.1468(45)   &  54713.513  & &  63099.2207(22)    &  63096.337  \\[0.5ex]
  $B$       & / MHz &   4862.362758(60) &   4858.512  & &   4766.557614(55)  &    4764.532  \\[0.5ex]
  $C$       & / MHz &   4458.249970(55) &   4455.474  & &   4425.560983(58)  &   4423.690  \\[0.5ex]
  $D_J$     & / kHz &      2.008283(40) &      2.021  & &      1.608429(49)  &      1.604  \\[0.5ex]
  $D_{JK}$  & / kHz &   -101.1809(21)   &   -103.4    & &   -108.8303(15)    &   -113.9    \\[0.5ex]
  $D_K$     & / MHz &      4.1783(39)   &     4.132   & &      6.19213(63)   &      6.452  \\[0.5ex]
  $d_1$     & / kHz &     -0.410926(21) &    -0.412   & &     -0.3010587(82) &     -0.301  \\[0.5ex]
  $d_2$     & / kHz &     -0.027954(24) &    -0.0239  & &     -0.0188128(25) &     -0.0154 \\[0.5ex]
  $H_{J}$   & / mHz     &  5.5695(83)   &     5.688   & &      4.035(13)     &      3.925  \\[0.5ex] 
  $H_{JK}$  & / Hz      & -0.4629(80)   &   -0.4464   & &     -0.46703(77)   &    -0.4788  \\[0.5ex] 
  $H_{KJ}$  & / Hz      & -6.733(28)    &   -7.879    & &     -3.690(47)     &    -3.192   \\[0.5ex]
  $H_{K}$   & / kHz     &  3.28(39)     &    0.7938   & &      1.042(46)     &     1.101   \\[0.5ex]
  $h_1$     & / mHz     &  2.1716(70)   &    2.189    & &      1.5502(23)    &     1.486   \\[0.5ex]
  $h_2$     & / mHz     &  0.3385(63)   &    0.2721   & &      0.2059(11)    &     0.165   \\[0.5ex]
  $h_3$     & / mHz     &  0.1202(45)   &    0.0733   & &      0.06093(23)   &     0.0450  \\[0.5ex]        
  $L_{JJK}$ & / $\mu$Hz &  3.06(38)     & \mcl{1}{c}{--} & &                 &  \mcl{1}{c}{--} \\[0.5ex]
  $L_{JK}$  & / mHz     & -0.3330(28)   & \mcl{1}{c}{--} & &  -0.3403(98)    &  \mcl{1}{c}{--} \\[0.5ex]
  $L_{KKJ}$ & / mHz     &               & \mcl{1}{c}{--} & &  -2.11(32)      &  \mcl{1}{c}{--} \\[0.5ex]
  $\chi_{aa}$             & / MHz & -4.0641(61) &  -4.1900  & &  1.035(39)  &  0.9311 \\[0.5ex]
  $\chi_{bb} - \chi_{cc}$ & / MHz & -2.654(15)  &  -2.7758  & & -7.6632(96) & -7.8950 \\[0.5ex]
  $\sigma_\mrm{w}$    &      &  0.91    &       & &  0.88   &        \\[0.5ex]
  \mcl{2}{l}{no. of lines}   &  \mcl{2}{c}{531} & &  \mcl{2}{c}{545} \\[0.5ex]
  \hline\hline \\[-2ex]
 \end{tabular}
 \end{center}
 $^a$ Equilibrium constants from extrapolated best structure, zero-point vibrational 
      corrections computed at fc-MP2/aug-cc-pVTZ. 
      Quartic centrifugal distortion constants computed using a composite scheme.
      Sextic centrifugal distortion constants computed at fc-CCSD(T)/cc-pVTZ.
      Nuclear quadrupole coupling constants computed at the ae-CCSD(T)/cc-pwCV5Z level. 
      See Appendix~\ref{sec:theor-details} for further explanation. \\
\end{table*}

\begin{table*}[h]
  \caption[]{Rotational, hyperfine, and vibrational partition functions for PGIM isomers.
             \label{tab:Qr}}
  \footnotesize
  \begin{center}
  \begin{tabular}{c c c D{.}{.}{3} c c c D{.}{.}{3}}
    \hline  \\[-1ex]
            & \mcl{3}{c}{$Z$-PGIM} & & \mcl{3}{c}{$E$-PGIM} \\[1ex]
    \cline{2-4} \cline{6-8} \\[-1ex]
    $T$ / K & \mcl{1}{c}{$Q_\text{rot}$} & \mcl{1}{c}{$Q_\text{HFS}$} & \mcl{1}{c}{$Q_\text{vib}$} &
            & \mcl{1}{c}{$Q_\text{rot}$} & \mcl{1}{c}{$Q_\text{HFS}$} & \mcl{1}{c}{$Q_\text{vib}$} \\[1ex]
    \hline  \\[-1ex]
      3  &    26.073   &     78.218  &  1.0000  & &    24.593   &     73.779  &  1.0000 \\[0.5ex]
      5  &    55.562   &     166.69  &  1.0000  & &    52.409   &     157.23  &  1.0000 \\[0.5ex]
     10  &    156.05   &     468.14  &  1.0000  & &    147.20   &     441.59  &  1.0000 \\[0.5ex]
     15  &    286.02   &     858.06  &  1.0000  & &    269.81   &     809.43  &  1.0000 \\[0.5ex]
     25  &    614.37   &    1843.11  &  1.0000  & &    579.58   &    1738.72  &  1.0000 \\[0.5ex]
     50  &   1736.09   &    5208.26  &  1.0026  & &   1637.91   &    4913.72  &  1.0022 \\[0.5ex]
    100  &   4911.56   &   14734.7   &  1.0674  & &   4634.74   &   13904.21  &  1.0637 \\[0.5ex]
    150  &   9028.77   &   27086.3   &  1.2375  & &   8520.22   &   25560.7   &  1.2301 \\[0.5ex]
    225  &  16602.6    &   49807.8   &  1.7010  & &  15634.8    &   46904.5   &  1.6883 \\[0.5ex]
    300  &  25569.7    &   76708.8   &  2.5137  & &  23925.1    &   71774.9   &  2.4962 \\
    \hline\hline \\[-2ex]
  \end{tabular}
  \end{center}
\end{table*}

The spectroscopic parameters for the two propargylimine isomers are reported in 
Table~\ref{tab:specpar} together with a compilation of theoretically computed 
quantities derived as described in Appendix~\ref{sec:theor-details} in the appendix.
Very precise determinations of the rotational constants $A,B,C$ were obtained, 
whose uncertainties are reduced by factors of 7--30 compared to the ones obtained in
the previous study \citep{Sugie-JMS85-PGIM}.
The quartic centrifugal distortion constants had also been determined in that work, 
but using an approximated perturbation expression \citep[described in][]{Watson-JCP67-CD}, 
so they are not directly comparable with the present results.
The precision of our determined values is very high: the standard errors are generally of 
a few 10$^{-3}$\%, only $D_K$ for the $Z$ isomer --- for which mainly $a$-type transition 
were recorded --- reaches 0.07\%\@.
A full set of sextic centrifugal distortion constants have been obtained for the first 
time, and their uncertainties are generally less than 1\% (12\% for $H_K$ of the $Z$ isomer).
Additionally, two octic centrifugal distortion constants ($L_{JJK}$, $L_{JK}$ for the 
$Z$ isomer and $L_{JK}$, $L_{JKK}$ for the $E$ isomer) had to be included in the fit
in order to reproduce all the measured transition frequencies within the experimental 
accuracy.
Their values should be considered as ``effective'' and they are determined with 15\% 
uncertainty at least.

The comparison between experimentally derived and theoretically computed values shows 
an overall excellent agreement.
For the rotational constants the deviations are generally lower than 0.1\%; 
the quartic centrifugal distortion coefficients show an average absolute deviation of 
less than 5\% for both isomers, with only the small $d_2$ constants showing larger 
deviations.
Also, the sextic centrifugal distortion constants compare well with the corresponding 
ab initio predictions; in all but a few cases, the deviations are within 10\%. 
Notable exceptions are the $H_K$ and $h_3$ values for the $Z$ isomer, for which only a few
$b$-type transitions have been recorded, thus making these parameter highly correlated with
the much larger $A$ and $D_K$ constants.

The theoretically computed values of the quadrupole coupling tensor components are also 
very close to the ones derived from the experimental hyperfine structure analysis.
In general, they agree within 3--4\%; only $\chi_{aa}$ for the $E$ isomer is slightly off
($\sim$ 10\%), but its experimental value is also affected by a larger uncertainty and it
is consistent with the ab initio prediction within $3\sigma$.

From the spectroscopic parameters presented in Table~\ref{tab:specpar} we have generated 
reliable sets of rest frequencies to guide astronomical searches of PGIM in the ISM.
To evaluate the overall precision of our spectral computations, we have chosen a subset
of transitions having $K_a = 0,1,2,3$ and lower level energy, $E/k < 200$\,K. 
Then, we further singled out all the lines having integrated intensity of at least 
$1/10$ of the maximum computed at 50\,K.
These selections constitute the most critical spectral data for an effective search of the 
PGIM isomers in astrophysical sources.
They contain 159~lines in the 54--265\,GHz range for $Z$-PGIM, and 247~lines in the 
66--551\,GHz range for the $E$-PGIM.
The maximum $1\sigma$ errors are 1.9\,kHz and 4.7\,kHz, respectively, which correspond to 
radial equivalent velocity uncertainties in the ranges $0.8-2.0\times 10^{-2}$ at 3\,mm, 
    and $0.3-0.8\times 10^{-2}$ at 1\,mm regime.

As electronic supplementary information, we provide a set of spectral catalogues
for PGIM isomers directly obtained with the SPCAT program 
\citep{Pick-JMS91-calpgm} without any further editing.
They exactly match the
CDMS\footnote{\texttt{https://cdms.astro.uni-koeln.de/cdms}} 
\citep{Muller-JMS05-CDMS,Endres-JMS16-CDMS} and 
JPL\footnote{\texttt{http://spec.jpl.nasa.gov/}} \citep{Pick-JQSRT98-JPL} file format, 
and are thus suited for a direct use in widespread astronomy line analysis  tools, 
such as CASSIS\footnote{%
   CASSIS is a software package for the analysis of astronomical spectra 
   developed by IRAP-UPS/CNRS (\texttt{http://cassis.irap.omp.eu}).} 
and MADCUBA \citep{Rivilla-ApJ16-PO}.
For each isomer, the results of two separate computations are provided. 
The files \texttt{x-pgim.cat} contain a listing of pure rotational frequencies 
extending up to 600\,GHz, whereas the files \texttt{x-pgim\_hfs.cat} provides a list of 
hyperfine components limited to 200\,GHz\@.
The first character (\texttt{x}) of the file names takes the values \texttt{z} or \texttt{e}
according to which isomer the file refers to.
In all catalogues, the integrated intensity of each transition is computed at 300\,K 
in order to comply with the CDMS standard.

A selection of rotational ($Q_\text{rot}$), hyperfine ($Q_\text{HFS}$), and vibrational 
($Q_\text{vib}$) partition functions for PGIM isomers is provided in Table~\ref{tab:Qr}.
The values are computed for temperatures ranging in the 3--300\,K interval, and they are 
obtained by direct summation over the rotational or hyperfine levels whose energy 
position is accurately determined during the spectral analysis.
The vibrational partition functions are computed through direct summation on all energy
levels (including combinations and overtones), which give a contribution to 
$Q_\text{vib}$ higher than $10^{-7}$\@.

\begin{table*}[t]
 \caption{Transitions of $Z$-PGIM detected towards G+0.693\@. 
          \label{tab:detec-lines}}
 \begin{center}
 \begin{tabular}{D{.}{.}{7} c cccc}
  \hline \\[-1ex]
  \mcl{1}{c}{Frequency$^a$} & Transition  & $\log I^{(b)}$ & $E_\mrm{up}$ & $ \int{T_{\rm A}^{*}} dv$ & Detection \\[0.5ex]
  \mcl{1}{c}{(GHz)}         &             &  (nm$^2$ MHz)  & (cm $^{-1}$) & (mK km s$^{-1}$)          & level$^c$ \\[1ex]
  \hline \\[-1ex]
   72.898955*  &  $8_{1,8} - 7_{1,7}$  & -4.315  & 10.18  &  281  &  8.0  \\[0.5ex]   
   74.355828*  &  $8_{0,8} - 7_{0,7}$  & -4.288  &  8.70  &  378  & 16.4  \\[0.5ex]   
   74.537358*  &  $8_{2,7} - 7_{2,6}$  & -4.328  & 15.37  &  107  &  4.2  \\[0.5ex]   
   74.742197*  &  $8_{2,6} - 7_{2,5}$  & -4.325  & 15.38  &  107  &  4.6  \\[0.5ex]   
   76.127838   &  $8_{1,7} - 7_{1,6}$  & -4.278  & 10.56  &  271  &  9.9  \\[0.5ex]   
   81.995559*  &  $9_{1,9} - 8_{1,8}$  & -4.166  & 12.61  &  222  &  8.0  \\[0.5ex]   
   83.587572   &  $9_{0,9} - 8_{0,8}$  & -4.141  & 11.18  &  295  & 13.5  \\[0.5ex]   
   83.842907*  &  $9_{2,8} - 8_{2,7}$  & -4.174  & 17.86  &   84  &  4.3  \\[0.5ex]   
   84.134809   &  $9_{2,7} - 8_{2,6}$  & -4.171  & 17.87  &   84  &  4.1  \\[0.5ex]   
   85.625901*  &  $9_{1,8} - 8_{1,7}$  & -4.129  & 13.10  &  209  &  5.4  \\[0.5ex]   
   91.086864*  & $10_{1,10}- 9_{1,9}$  & -4.034  & 15.35  &  162  & 10.1  \\[0.5ex]   
   92.797807*  & $10_{0,10}- 9_{0,9}$  & -4.010  & 13.96  &  212  & 14.1  \\[0.5ex]   
   95.117382*  & $10_{1,9} - 9_{1,8}$  & -3.997  & 15.95  &  149  &  7.0  \\[0.5ex]   
  100.172426*  & $11_{1,11}-10_{1,10}$ & -3.916  & 18.39  &  109  &  3.0  \\[0.5ex]   
  101.984662   & $11_{0,11}-10_{0,10}$ & -3.894  & 17.06  &  141  &  4.7  \\[0.5ex]   
  104.601405*  & $11_{1,10}-10_{1,9}$  & -3.880  & 19.13  &   98  &  3.0  \\[0.5ex]   
  109.251842   & $12_{1,12}-11_{1,11}$ & -3.809  & 21.73  &   68  &  3.2  \\[0.5ex]   
  111.146539   & $12_{0,12}-11_{0,11}$ & -3.789  & 20.46  &   87  &  2.7  \\[0.5ex]   
  \hline\hline \\[-2ex]
 \end{tabular}
 \end{center}
 $^a$ Asterisks denote the transitions used in the MADCUBA$-$AUTOFIT analysis. \\
 $^b$ base~10 logarithm of the integrated intensity of the transition at 300\,K\@. \\
 $^c$ See text of Sect.~\ref{sec:obs-res}.
 \vspace{0.5cm}
\end{table*}

\begin{table*}[h]
 \caption{Derived physical parameters for PGIM isomers and proposed molecular precursors. 
          \label{tab:obs-param}}
 \begin{center}
 \begin{tabular}{l cccccc}
  \hline \\[-1ex]
  Molecule   & $N$                    & $T_\mrm{ex}$  & $v_\mrm{LSR}$  & FWHM           & Abundance$^a$        & Reference$^b$ \\[0.5ex]
             & ($10^{14}$\,cm$^{-2}$) & (K)           & (km\,s$^{-1}$) & (km\,s$^{-1}$) & ($\times$10$^{-10}$) &               \\[1.5ex]
  \hline \\[-1ex]
  $Z$-PGIM         &  $0.24 \pm 0.02$ &  8            &  69           & 20        &    1.8  & 1 \\[0.5ex]
  $E$-PGIM         &  $<0.13$         &  8            &  69           & 20        & $<0.9$  & 1 \\[1.0ex]
  \ce{CH3CCH}$^c$  &  $17.0\pm 2$     & $19\pm 1$     &  69           & 20        &   126   & 1 \\[0.5ex]
  \ce{HC3N}$^d$    &  $7.1\pm 1.3$    & $12\pm 2$     &  $68\pm 1$    & $22\pm 1$ &    53   & 2 \\[0.5ex]
  \ce{CH2CHCN}     &  $0.9\pm 0.1$    & $10.8\pm 1.1$ &  $68\pm 1$    & $22\pm 2$ &     7   & 2 \\[0.5ex]
  \ce{CCH}$^d$     &  $53\pm 7$       & --            &  --           & --        &   391   & 1 \\[0.5ex]
  $Z$-\ce{NCCHNH}  &  $2.0\pm 0.6$    & $8\pm 2$      & $68.3\pm 0.8$ & 20        &    15   & 3 \\[0.5ex]
  $E$-\ce{NCCHNH}  &  $0.33\pm 0.03$  & 8             & $68.0\pm 0.8$ & 21$\pm$ 2 &   2.4   & 3 \\[0.5ex]
  \ce{HCCCHO}      &  $0.32 \pm 0.02$ & $18\pm 2$     & $67.6\pm 0.4$ & 20        &   2.4   & 4 \\[0.5ex]
  \hline\hline \\[-2ex]
 \end{tabular}
 \end{center}
 $^a$ We adopted $N_{\mrm{H}_2} = 1.35\times 10^{23}$\,cm$^{-2}$ as inferred by \citet{Martin-ApJ08-shocks} 
      from C$^{18}$O observations. \\
 $^b$ References: (1) This work; (2) \citet{Zeng-MNRAS18-COMs}; (3) \citet{Rivilla-MNRAS19-CNMtIm}; (4) \citet{Rivilla-inprep}\@. \\
 $^c$ Fit obtained from the $J=9-8$, $8-7$, $6-5$ and $5-4$ transitions with $K=0$ and $K=1$. \\
 $^d$ The column densities of \ce{HC3N} and \ce{CCH} were derived from the optically thin transitions
      of the \ce{H^{13}CCCN} and \ce{C^{13}CH} isotopologues, respectively. 
      The isotopic ratio of $^{12}\mrm{C}/^{13}\mrm{C}\sim 21$ measured by \citet{ArmAben-MNRAS15-GC} 
      in G+0.693 was used. \\
\end{table*}

\section{Detection of PGIM in the Galactic Center cloud G+0.693} \label{sec:detect}

\subsection{Observations} \label{sec:observ}
\indent\indent
We have searched for PGIM towards the molecular cloud G+0.693-0.027 (G+0.693 hereafter) 
located in the ``Central Molecular Zone'' (CMZ), the inner $\sim$500\,pc of our Galaxy. 
G+0.693, located at $\sim$1\,arcmin north-east of the star-forming protocluster SgrB2(N), 
does not show any sign of on-going massive star formation such as \ce{H2O} masers, 
\ion{H}{ii} regions or dust continuum sources \citep[e.g.,][]{Ginsburg-ApJ18-SgrB2}. 
However, despite being a quiescent cloud, it is one of the main repositories of COMs in 
the Galaxy \citep{ReqTorr-ApJ08-OCOM,Zeng-MNRAS18-COMs}.
Among the many molecules detected in this cloud, there are several species directly related 
with prebiotic chemistry like the simplest sugar glycolaldehyde 
\citep[\ce{CH2OHCHO};][]{ReqTorr-AA06-HCC}, formamide and methyl isocyanate (\ce{NH2CHO} 
and \ce{CH3NCO}, respectively; \citealt{Zeng-MNRAS18-COMs}), phosphorous-bearing species 
such as PO \citep{Rivilla-MNRAS18-Pmol}, and recently urea 
\citep[\ce{NH2CONH2},][]{JimSerra-inpress-Urea}\@.
In addition, several imines have been reported towards G+0.693: 
methanimine \citep[\ce{CH2NH},][]{Zeng-MNRAS18-COMs},
ethanimine \citep[\ce{CH3CHNH},][]{Rivilla-inprep}, and the $E,Z$ isomers of 
$C$-cyanomethanimine, \ce{NCCHNH}, \ce{NCCHNH} \citep{Rivilla-MNRAS19-CNMtIm}, a possible 
precursor of adenine, which is one of the DNA and RNA nucleobases.
For $Z$-\ce{NCCHNH}, it has been the first detection in the ISM.
Therefore, G+0.693 is a promising target for the detection of new imines such as PGIM. 

We have searched for PGIM in a spectral survey of G+0.693 conducted with the IRAM~30\,m 
telescope at 3~and 2\,mm. 
The observations were performed in two different observing runs during 2019: 
April 10-16 (project 172-18), and August 13-19 (project 018-19).
We used the broad-band Eight MIxer Receiver (EMIR) and the fast Fourier transform spectrometers 
in FTS200 mode, which provided a channel width of $\sim$200\,kHz, i.e. a velocity resolution 
of $\sim$0.35--0.85\,km\,s$^{-1}$. 
Since we are interested in weak line emissions, we smoothed  all the 
spectra to 5\,km\,s$^{-1}$ velocity resolution, enough to resolve the line widths of 
$\sim$20\,km\,s$^{-1}$ measured in this source. 
Each spectral setup was observed several times slightly changing the central frequency 
(by 20--100\,MHz shifts) in order to identify possible spectral features resulting from 
unsuppressed image side band emission.

\begin{figure*}[t]
 \includegraphics[width=0.8\textwidth]{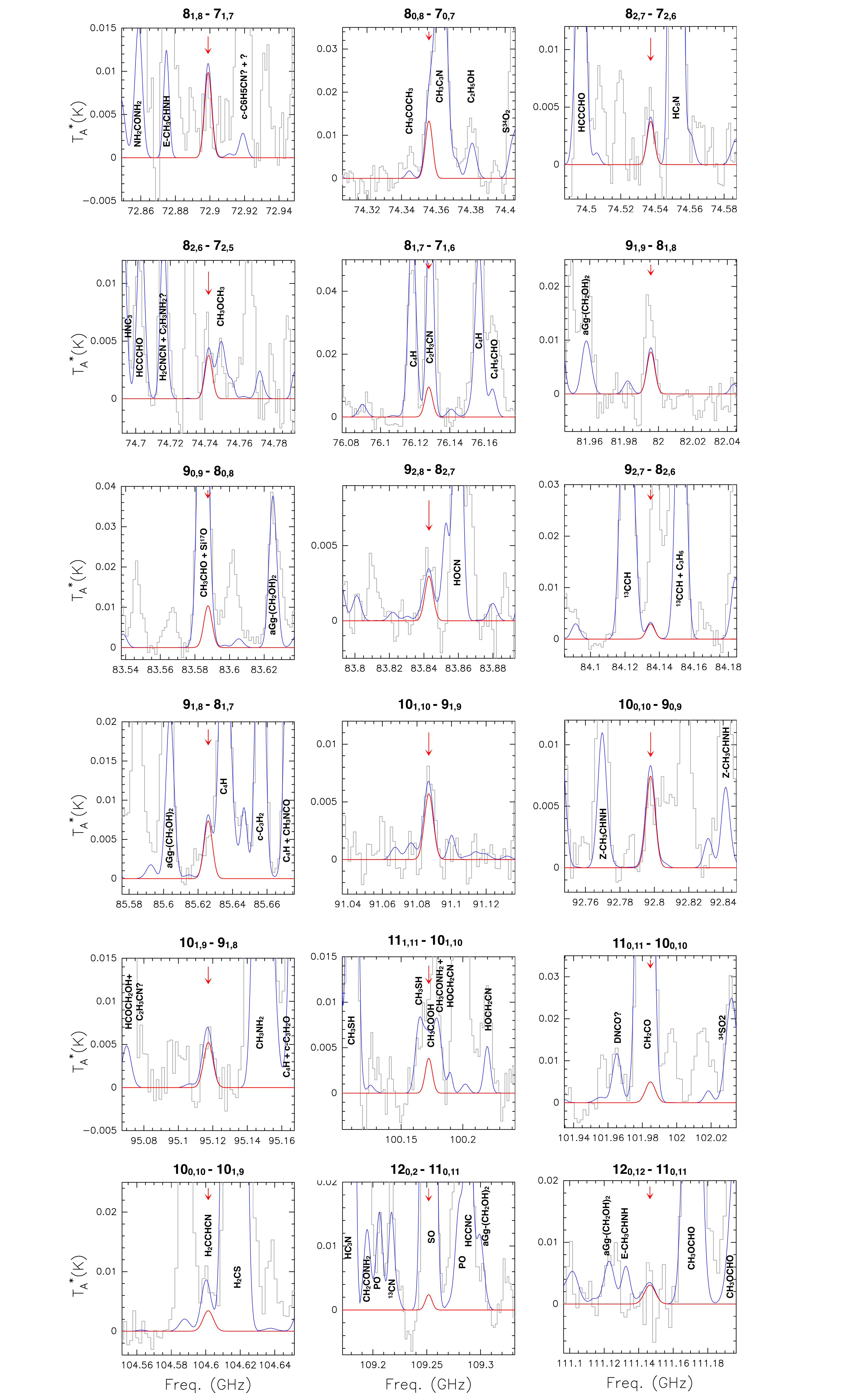}
 \centering
 \caption{LTE spectrum (in red) of the 18 brightest transitions of $Z$-PGIM detected 
          towards G+0.693. 
          The contribution of all the species identified in the source so far is indicated 
          with a blue curve.
          \label{fig:detec-spectra}}
\end{figure*}

\subsection{Data analysis and results} \label{sec:obs-res}
\indent\indent
The two intervals of frequency covered by the survey are 71.76--116.72\,GHz 
and 124.77--175.5\,GHz\@.
During the observations, the pointing was checked every 1--1.5\,h on nearby planets, 
QSOs or bright \ion{H}{ii} regions, and the telescope focus was checked at the 
beginning of the observations and after sunset and sunrise. 
The half-power beam width (HPBW) of the telescope at the observed frequencies is 
14\arcsec--33\arcsec.
The observations were centred at the coordinates of G+0.693: 
$\alpha$(J2000) = \hms{17}{47}{22} and $\delta$(J2000)= \dms{-28}{21}{27}\@.
The position switching mode was used in all the observations, with an off position of 
(-885\arcsec, 290\arcsec)\@.
The line intensity of the spectra is given in $T_\mrm{A}^*$ as the molecular emission 
towards G+0.693 is extended over the full beam
\citep{ReqTorr-AA06-HCC,Martin-ApJ08-shocks,Rivilla-MNRAS18-Pmol}.
Contamination by image-band lines have been identified and eliminated during the data 
reduction comparing the same frequency band observed with two different spectral 
setups.

The identification of the molecular lines was performed using the new 
spectroscopic data of PGIM and the SLIM (Spectral Line 
Identification and Modelling) tool within the MADCUBA package\footnote{
  Madrid Data Cube Analysis on ImageJ is a software developed at the Centre of Astrobiology 
  (CAB) in Madrid; \texttt{http://cab.inta-csic.es/madcuba/Portada.html}.
} 
\citep{Martin-AA19-SLIM}.
SLIM generates synthetic spectra of molecular species under the assumption of Local 
Thermodynamic Equilibrium (LTE) conditions.
We used the MADCUBA-AUTOFIT tool that compares the observed spectra with LTE synthetic 
spectra, and provides the best non-linear least-squares fit using the Levenberg-Marquardt 
algorithm \citep[see details in][]{Martin-AA19-SLIM}.
The free parameters of the fit are: the molecular column density ($N$), the excitation 
temperature ($T_\mrm{ex}$), the systemic velocity ($v_\mrm{LSR}$) and the full width 
half maximum (FWHM) of the line Gaussian profiles. 
Since the fit convergence was difficult to achieve when trying to optimise all four 
parameters, we fixed the $v_\mrm{LSR}$ and the FWHM to 69\,km\,s$^{-1}$ and 20\,km\,s$^{-1}$, 
respectively.
These values allowed to reproduce well all the observed spectra and are consistent with 
the ones derived for other imines in the same source
 \citep{Zeng-MNRAS18-COMs,Rivilla-MNRAS19-CNMtIm}.
We have also fixed the excitation temperature to 8\,K, as found for $C$-cyanomethanimine  
\citep{Rivilla-MNRAS19-CNMtIm} as well as for other complex organic molecules 
\citep{ReqTorr-AA06-HCC,ReqTorr-ApJ08-OCOM}.
This value also compares well to that obtained for the simplest imine, \ce{CH2NH}, 
which was 9.7$\pm$0.4\,K \citep{Zeng-MNRAS18-COMs}.

The 18~brightest transitions of $Z$-PGIM detected in our data at level $>2.5\sigma$ are 
illustrated in Fig.~\ref{fig:detec-spectra}. 
The resulting line parameters are summarised in Table~\ref{tab:detec-lines}.
For each transition, we have evaluated the possible blending with other molecular species. 
We have searched in the spectral survey for more than 300~different molecules, which 
include all the species detected so far in the ISM\footnote{
  \texttt{https://cdms.astro.uni-koeln.de/classic/molecules}.
}.
The complete list of the molecular species detected in the survey will be presented in a 
forthcoming paper.
For the scope of this work, we show in Fig.~\ref{fig:detec-spectra} the contribution 
of the molecules identified in the spectral intervals around the transitions assigned 
for $Z$-PGIM. 
Note that the low excitation temperature of the COMs detected in this source guarantees 
that only the lowest energy levels of these molecules are populated 
(see Table \ref{tab:detec-lines}). 
This implies that line confusion is expected to be less severe in this source than in 
hot core sources such as SgrB2(N) \citep[e.g.,][]{Belloche-AA19-ReMoCA}
in spite of the larger line widths. 
Therefore, the level of line blending of the molecular lines, especially at 3\,mm, 
is expected to be low.

We have run AUTOFIT to derive the column density of $Z$-PGIM using the transitions that 
are less blended with other species (indicated with an asterisk in 
Table~\ref{tab:detec-lines}). 
The resulting simulated spectra are presented in red solid lines in 
Fig.~\ref{fig:detec-spectra}.
The $Z$-PGIM transitions have integrated intensities 
$\int{T_\mrm{A}^*\mrm{d}v} > 68$\,mK\,km\, s$^{-1}$ (see Table~\ref{tab:detec-lines}).
We calculated the detection level of each transition by comparing the velocity-integrated 
intensity with $\sigma = \text{rms}\times\sqrt{\delta v/\text{FWHM}}\times\text{FWHM}$, 
where $\text{rms}$ is the noise root-mean-square measured over a line-free spectral range 
of $\pm$500\,km\,s$^{-1}$ around each transition, and $\delta v$ is the spectral resolution 
expressed in velocity units. 
The derived $\text{rms}$ are in the range 1.5--3.9\,mK\,km\, s$^{-1}$ in channels 
of~5\,km\,s$^{-1}$\@.
As indicated in Table~\ref{tab:detec-lines}, 14~transitions are detected above $4\sigma$, 
and 8~transitions are above $6\sigma$\@. 
The integrated intensities corresponding to the fitted column densities derived by AUTOFIT 
are shown in Table~\ref{tab:obs-param}. 
The column density of $Z$-PGIM is $(0.24\pm 0.02)\times 10^{14}$\,cm$^{-2}$\@. 
By considering the \ce{H2} column density towards G+0.693, 
$N_{\mrm{H}_2}=1.35\times 10^{23}$\,cm$^{-2}$ \citep{Martin-ApJ08-shocks}, 
we can derive a $Z$-PGIM fractional abundance of $1.8\times 10^{-10}$\@.

The higher-energy $E$-PGIM isomer is not detected in the data. 
To derive its upper limit we have used the two brightest spectral features that appear 
completely free of contamination by other species in the observed spectra: 
the hyperfine $\Delta F=+1$ triplet of the $J_{K_a,K_c} = 5_{1,5}-4_{0,4}$ 
line centred at 101\,915\,MHz ($E_\mrm{L}=3.07$\,cm$^{-1}$), and the one of the 
$J_{K_a,K_c} = 6_{1,6}-5_{0,5}$ line at 110\,103\,GHz ($E_\mrm{L}=4.60$\,cm$^{-1}$).
We adopted the same assumptions on excitation temperature, velocity and line FWHM as for 
the $Z$-PGIM.
The upper limits to the integrated intensity are derived using the formula 
$3\times\text{rms}\times\Delta v/\sqrt{n_\text{chan}}$, where $n_\text{chan}$ is the 
number of channels covered by the full line width 
$\Delta v$.
The upper limit derived for the column density of $E$-PGIM is 
$N<1.3\times 10^{13}$\,cm$^{-2}$\@.
This implies an isomer abundance ratio of $[Z]/[E]>1.9$, hinting at a higher abundance
of the thermodynamically more stable form.

We also report fractional abundances of other molecular species that might be 
chemically related to PGIM: cyanoacetylene (\ce{HC3N}), acrylonitrile (\ce{CH2CHCN}), 
propyne (\ce{CH3CCH}), $C$-cyanomethanimine (\ce{NCCHNH}), ethynyl (\ce{CCH}),
and propynal (\ce{HCCCHO}).
All of them are detected towards G+0.693 and have abundances $\gtrsim 10^{-9}$.
The derived parameters are presented in Table~\ref{tab:obs-param}.
For \ce{CH3CCH}, we have fitted the $J=9-8$, $8-7$, $6-5$ and $5-4$ transitions with 
$K=0,1$. 
For \ce{CCH}, we derived its abundance by fitting the optically thin lines of the 
isotopic variant \ce{C^{13}CH}, and assuming the isotopic ratio of 
\ce{^{12}C}/\ce{^{13}C}$\sim 21$ measured in G+0.693 by \citet{ArmAben-MNRAS15-GC}.
For \ce{HC3N}, \ce{CH2CHCN}, and \ce{NCCHNH}, we adopted the values obtained in previous 
works \citep{Zeng-MNRAS18-COMs,Rivilla-MNRAS19-CNMtIm}, whereas the \ce{HCCCHO} results
are taken from a full chemical analysis of the source that will be presented in a  
forthcoming paper \citep{Rivilla-inprep}. 

\begin{figure*}[t]
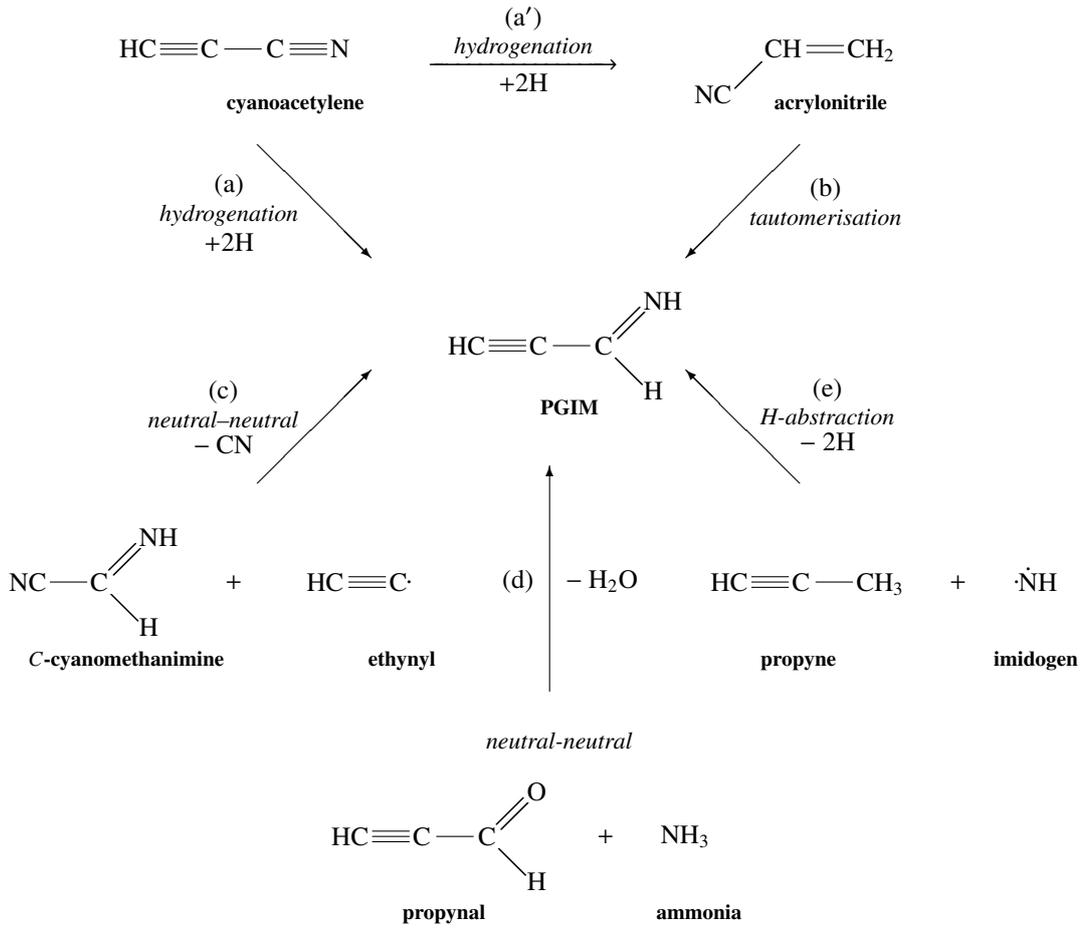

 \centering
 \begin{minipage}{\textwidth}
 \begin{tabular}{c c c c c}
 %
 \mcl{2}{c}{\hspace{10pt}
            \cdonecell{0pt}{3.5cm}{\hspace{-15pt}
             \tetrahedral{0==C;2T==HC;%
                          4==\tetrahedral{0==C;2==(yl);4T==N}}}}
 &  
 \hspace{-30pt}\reactrarrow{0pt}{2.5cm}{\shortstack{(a$^\prime$)\\{\footnotesize\it hydrogenation}}}{+2H}  
 & 
 \mcl{2}{c}{\hspace{-70pt}
            \cdonecell{0pt}{3.5cm}{\hspace{-15pt}
             \ltrigonal{0==CH;1D==CH$_2$;3==NC}}}
           
 \\[-15pt]
 {\tiny\bf\hspace{90pt} cyanoacetylene \hspace{-80pt}} 
 & & & & 
 {\tiny\bf\hspace{-95pt} acrylonitrile} 
 \\[10pt]
 %
 & 
 \reactsearrow{0pt}{1.5cm}{}{\llap{\raisebox{-10pt}{
                                   \shortstack{(a)                             \\ 
                                              {\footnotesize\it hydrogenation} \\ 
                                               +2H}\hspace{0pt}}}} 
 & & 
 \reactswarrow{0pt}{1.5cm}{}{\rlap{\hspace{-6pt}
                                   \shortstack{(b)                             \\
                                               \footnotesize\it tautomerisation}}}
 \\[10pt]
 %
 \mcl{5}{c}{\hspace{5pt}
            \makebox(0,400){%
             \tetrahedral{0==C;2T==HC;%
                          4==\rtrigonal{0==C;1==(yl);2==H;3D==NH}}}}
 \\[-5pt]
 \mcl{5}{c}{\tiny\bf\hspace{30pt}PGIM} \\[-20pt]
 %
 & 
 \reactnearrow{0pt}{1.5cm}{\llap{\raisebox{-10pt}{
                                  \shortstack{(c)                                 \\ 
                                              {\footnotesize\it neutral--neutral} \\ 
                                               $-$\:\ce{CN}}\hspace{0pt}}}}{} 
 & &
 \reactnwarrow{0pt}{1.5cm}{}{\rlap{\hspace{10pt}
                                   \shortstack{(e)                              \\ 
                                               {\footnotesize\it H-abstraction} \\ 
                                               $-$\:2H}}}
 & \\[-10pt]
 %
 \mcl{2}{c}{\hspace{-60pt}
            \cdonecell{0pt}{2.4cm}{%
             \rtrigonal{0==C;1==NC;2==H;3D==NH}} \quad + \quad 
             \cdonecell{0pt}{1.8cm}{%
              \hspace{-20pt}\tetrahedral{0==HC;4T=={\hspace{0pt}}\chemradicalA[2]{C}}}}\hspace{-60pt} 
 &
 \hspace{-15pt}\reactuarrow{0pt}{3.0cm}{\hspace{20pt}(d)}{$-$\:H$_2$O\hspace{-20pt}} 
 &
 \mcl{2}{c}{\cdonecell{0pt}{3cm}{%
             \tetrahedral{0==C;2T==HC;4==CH$_{3}$}} 
             \quad + \quad \cdonecell{0pt}{1cm}{\chemradicalA[14]{N}\hspace{0pt}H}}
 \\[-20pt]
 \mcl{2}{c}{\tiny\bf\hspace{16pt} $C$-cyanomethanimine \hspace{50pt} ethynyl \hspace{10pt}} 
 & & 
 \mcl{2}{c}{\tiny\bf\hspace{26pt} propyne \hspace{55pt} imidogen} \\[20pt]
 %
 \mcl{5}{c}{\footnotesize\it\hspace{20pt} neutral-neutral} \\[10pt]
 \mcl{5}{c}{\makebox(0,400){%
             \cdonecell{0pt}{3cm}{%
              \hspace{-20pt}\tetrahedral{0==C;2T==HC;%
                                         4==\rtrigonal{0==C;1==(yl);2==H;3D==O}}} 
             \quad + \quad \cdonecell{0pt}{1cm}{NH$_3$}}
           } \\                                   
 \mcl{5}{c}{\tiny\bf\hspace{30pt} propynal \hspace{60pt} ammonia} \\ 
 \end{tabular}
 \end{minipage}
 \caption{Possible formation mechanisms of PGIM in the ISM.
          \label{fig:react}}
\end{figure*}

\subsection{Discussion} \label{sec:form-routes}
\indent\indent
To the best of our knowledge, the formation of PGIM in the ISM has never been
investigated to date, thus there are no chemical pathways included in astrochemical 
databases such as KIDA\footnote{
  \texttt{http://kida.obs.u-bordeaux1.fr}\@.
} \citep{Wakelam-ApJ12-KIDA} or UMIST\footnote{
  \texttt{http://udfa.a jmarkwick.net/index.php}\@.
} \citep{McElroy-AA13-UMIST}. 
In the following, we formulate and discuss a few possible PGIM formation mechanism 
based on the results of theoretical and experimental studies on detected imines 
\citep[e.g.][]{Lovas-ApJ06-KetIm,Theule-AA11-hydHCN,Krim-MNRAS19-MKetIm}:
\begin{enumerate}[label=(\alph*)]
 \item hydrogenation of cyanoacetylene (possibly on dust grains);
 \vspace{0.5ex}
 \item tautomerisation of acrylonitrile;
 \vspace{0.5ex}
 \item neutral--neutral reaction between $C$-cyanomethanimine and the ethynyl radical 
       to form PGIM and the CN radical;
 \vspace{0.5ex}
 \item reaction of propynal and ammonia followed by water elimination; 
 \vspace{0.5ex}
 \item hydrogen abstraction on propyne which gives PGIM after reaction with the 
       imidogen radical.
\end{enumerate}
The schematic reactions are illustrated in Fig.~\ref{fig:react}.
As shown in Table \ref{tab:obs-param}, all the proposed processes involve
molecular precursors that are significantly more abundant than PGIM
(although propynal only marginally), therefore it is not unlikely that one (or several) 
of these mechanisms might be able to account for the presence of this imine in G+0.693.

The route (a) involves selective hydrogenation of the \ce{C#N} group on dust grain surfaces.
This process has proved to be effective for generating fully saturated methylamine 
from HCN ices \citep{Theule-AA11-hydHCN}, although the intermediate imine was not observed.
More doubtful is the viability of this chemical scheme starting from larger nitriles.
For example, the formation of ethanimine from \ce{CH3CN}, as suggested by 
\citet{Loomis-ApJ13-EthIm}, seems to be unlikely in view of recent laboratory works 
\citep{Nguyem-AA19-amines}, which investigated the co-deposition of methyl 
cyanide and H atoms and showed that the \ce{C#N} moiety is not reduced in the 10--60\,K 
temperature range.
In the absence of energetic processes, the hydrogenation of cyanocetylene would result 
in a competition of the H attack to the \ce{C#C} and \ce{C#N} triple bonds, leading mainly 
to acrylonitrile (\citealt{Krim-MNRAS19-MKetIm}, route a$^\prime$). 
However, route (a) cannot be completely ruled out as the chemistry in G+0.693 is known 
to be shock-dominated \citep[e.g.,][]{Rivilla-MNRAS19-CNMtIm}.
Also, cosmic-ray ionisation rate is expected to be high across the Galactic Centre
\citep{Goto-JPCA13-H3+} and also likely high in G+0.693 \citep[see][]{Zeng-MNRAS18-COMs},
thus providing an additional source of energy.

Route (b) is analogous to the proposed \ce{CH3CN}$\rightarrow$\ce{CH2=C=NH} conversion 
driven by shocks \citep{Lovas-ApJ06-KetIm}.
The energetics of the $Z$-PGIM$\rightarrow$\ce{CH2=CHCN} tautomerisation has been 
investigated theoretically by \citet{Osman-IJMS14-PGIM}. 
Their calculations (at the MP2 level of theory) show that acrylonitrile is more stable than
$Z$-PGIM by 32.7\,kcal\,mol$^{-1}$ ($E/k\approx 16\,500$\,K), and the double H migration 
involved in the process has barrier energies exceeding 80\,kcal\,mol$^{-1}$ 
($E/k\approx 40\,300$\,K)\@.
They concluded that \ce{CH2=CHCN} is dominant in the cold ISM, and point to a conversion 
to PGIM possibly occurring in shock-dominated regions or hot cores.
The above figures however, seems to be too high to allow for such a pathway in the shocks 
affecting G+0.693 \citep{ReqTorr-AA06-HCC}.

Route (c), (d), and (e), require two molecules to be co-spatial.
In all these processes, at least one reactant is much more abundant than PGIM in G+0.693: 
(ethynyl, propyne, and ammonia), hence they are feasible in principle.
Route (d) seems to be less likely because propynal is only marginally more abundant than 
PGIM, unlike the other two reactants, $C$-cyanomethanimine and propyne, which are factors 
of $\sim 10$ and $\sim 100$ more abundant, respectively. 
However, given the lack of information on the associated reaction rates and energy 
barriers, only speculative reasoning can be done.

In analogous cases, hints to constrain the formation scenario can be provided 
by the observed $[Z]/[E]$ isomer abundance ratio.
It has been proposed that the relative abundances of structural isomers in the ISM 
might be established by their thermodynamic stabilities 
\citep[minimum energy principle, see][]{Latte-ApJL09-iCOMs}.
Although there are some well known exceptions 
(H$_4$C$_2$O$_2$ and H$_2$C$_3$O isomers; see also \citealt{Shingl-ApJ2019-H2C3O}),
this hypothesis seems to work well, at least for simple cases of geometrical isomerism.
\citet{Rivilla-MNRAS19-CNMtIm}, for example, showed that the ratio between the $Z$- and 
$E$- isomers of $C$-cyanomethanimine (\ce{NCCHNH}) does follow thermodynamic equilibrium:
\begin{equation} \label{eq:equil}
 [Z]/[E] = \frac{N(Z)}{N(E)} = 
                    \frac{1}{g}\exp\left(\frac{\Delta E}{T_\mrm{k}}\right) \,,
\end{equation}
where $\Delta E$ is the energy difference between the two isomers, $T_\mrm{k}$ is the 
kinetic temperature of the gas, and $g$ is a factor that accounts for statistical 
weights (1 for the $Z$- and $E$- isomers of both \ce{NCCHNH} and PGIM).
The $[Z]/[E]\sim 6$ found for $C$-cyanomethanimine, implies a $T_\mrm{k}$ 
in the 130--210\,K range, which is in good agreement with the kinetic temperature 
measured by \citet{Zeng-MNRAS18-COMs} in G+0.693\footnote{
  The $T_\mrm{k}$ value found for G+0693 is significantly lower than the average 
  $T_\mrm{ex}$ of the observed molecules (see Table~\ref{tab:obs-param}) as, due 
  to the low density of the source, their rotational energy level manifolds are 
  sub-thermally populated.
}\@.

According to our newly performed ab initio calculations (see Appendix~\ref{sec:theor-details}), 
the $\Delta E/k$ for PGIM isomers is 428.2\,K\@. 
Using this value, the $[Z]/[E]$ ratio expected for PGIM for the $T_\mrm{k}$ range 
130--210\,K, is 8--27\@.
This prediction is consistent with the observed lower limit of $[Z]/[E] > 1.9$, but 
the lack of a more stringent upper limit for $E$-PGIM prevents a firm confirmation
that PGIM isomers actually follow the relative abundance predicted by thermodynamics.

The cosmic-ray ionization rate in the CMZ where G+0.693 is located is expected to be 
factors of 10--100 higher in the Galactic centre than in the disc, as measured by 
\citet{Goto-JPCA13-H3+,Goto-ApJ14-H3+} using \ce{H3+} observations. 
\citet{Zeng-MNRAS18-COMs} suggested that a relatively high cosmic-ray ionisation 
rate of $1-10\times 10^{-15}$\,s$^{-1}$ might be responsible of the cyanopolyynes and 
nitriles molecular ratios found in G+0.693\@.
Indeed, this produce an enhanced abundance of C atoms, due to the efficient CO 
destruction \citep{Bisbas-MNRAS19-Simul}, possibly accounting for the large abundance of 
carbon-chains in this region. 
Also, a high cosmic-ray ionisation rate can create atomic hydrogen (H) in an efficient 
way from the dissociation of \ce{H2} \citep{Padov-AA18-CR}; then, different reactivity 
of the PGIM isomers with atomic H might produce the different abundances observed. 
However, theoretical quantum chemical calculations of the reactions of PGIM isomers 
with H need to be performed to test this possibility. 

Another possible destruction route is through charged species such as \ce{H3+}. 
In this case, the destruction rates depend on the permanent dipole moment of the 
different isomers.
For $C$-cyanomethanimine, the $E$ isomer presents a dipole moment that is a factor of 
$\sim 3$ higher than the one of the $Z$ isomer. 
\citet{Shingl-submit} has shown theoretically that this produces a more efficient 
destruction of the $E$ by reaction with \ce{H3+}, which may contribute to its lower 
abundance observed in G+0.693. 
This explanation, however, cannot be applied to PGIM, since both $Z$ and $E$ isomers
have very similar modulus of their permanent dipole moments (see Section~\ref{sec:mol-prop}). 
Therefore, one should expect similar destruction rates for both isomers with \ce{H3^+}.

\section{Conclusions} \label{sec:conc}
\indent\indent
This paper presents an extensive theoretical and laboratory study of the rotational 
spectrum of PGIM in its vibrational ground state, extending the earlier, very limited 
knowledge on the spectroscopic properties of this simple imine.
The recordings have been performed in selected frequency intervals spanning the 
83--500\,GHz range, collecting some 500~lines for each of the two $E$ and $Z$ 
geometrical isomers.
These experimental data were fitted to the coefficients of the $S$-reduced rotational 
Hamiltonian, providing a very precise set of rotational, quartic, and sextic centrifugal 
distortion constants.
Many transitions, especially those of $b$-type dipole, show a resolvable hyperfine patterns 
due to the quadrupole coupling of the \ce{^{14}N} nuclei.
For these transitions, 357~separate components were accurately measured and analysed
to determine the corresponding hyperfine coupling coefficients.
The optimised values of all the spectroscopic constants are in excellent agreement 
with the results of high-level theoretical calculations, which were performed to 
assist the analysis of the laboratory data.

The newly obtained set of spectroscopic parameters (Tables~\ref{tab:specpar}) allowed 
to generate a highly precise set of rest-frequencies for $E$- and $Z$-PGIM at mm regime.
With these data we have searched for PGIM in a spectral survey of the molecular cloud 
G+0.693 located in the ``Central Molecular Zone''.
We have detected 18~transitions of $Z$-PGIM, the lowest energy isomer, for which a column 
density of $N = (0.24\pm 0.02)\times10^{14}$\,cm$^{-2}$ was derived.
The higher-energy $E$-PGIM isomer was not detected in the data, setting an upper limit of 
$N < 1.3\times 10^{13}$\,cm$^{-2}$ from the two strong spectral features 
which show no contamination from other species.

The fractional abundance (w.r.t.~\ce{H2}) derived for $Z$-PGIM is $1.8\times 10^{-10}$.
This value was compared with the ones found for possible chemical precursors, i.e., 
cyanoacetylene (\ce{HC3N}), propyne (\ce{CH3CCH}), acrylonitrile (\ce{CH2CHCN}), 
$C$-cyanomethanimine (\ce{NCCHNH}), ethynyl (\ce{CCH}), and propynal (\ce{HCCCHO}). 
The relative abundance of all of them detected towards G+0.693 are higher 
(by up to two orders of magnitude for \ce{CH3CCH}) than that of PGIM\@.

\begin{acknowledgement} \label{sec:ack}
We thank the IRAM-30m staff for the precious help during the different observing runs.
A.P.C.  gratefully acknowledges financial support by University C\`a Foscari Venezia 
(ADiR funds) and the super-computing facilities of CINECA (project ``CINEMA'', grant HP10C2QE6F) 
and SCSCF (``Sistema per il Calcolo Scientifico di C\`a Foscari'', a multiprocessor cluster 
system owned by Universit\`a C\`a Foscari Venezia).
V.M.R. has received funding from the European Union's Horizon 2020 research and innovation 
programme under the Marie Sk\l{}odowska-Curie grant agreement No 664931. 
I.J.-S. and J.M.-P. have received partial support from the Spanish FEDER 
(project number ESP2017-86582-C4-1-R), and State Research Agency (AEI) through project 
number MDM-2017-0737 Unidad de Excelencia Mar\'ia de Maeztu$-$Centro de Astrobiolog\'ia 
(INTA-CSIC).
J.C.G. thanks the program Physique et Chimie du Milieu Interstellaire (INSU-CNRS), 
the Centre National d'Etudes Spatiales (CNES) and the PHC Procope 42516WC (CAMPUS France) 
for financial support.
D.P. and L.B and acknowledge DAAD support through the project 57445281 ``Small organic 
molecules in the Interstellar Medium'' in the framework of ``Programm Projektbezogener 
Personenaustausch Frankreich (Phase I) 2019''.
S.Z. acknowledges support from RIKEN Special Postdoctoral Researcher Program. 
\end{acknowledgement}

\appendix
\section{Details of the theoretical calculations} \label{sec:theor-details}
\indent\indent
For calculating the equilibrium structures, extrapolation to the complete basis set (CBS) 
limit was performed for both the Hartree-Fock self-consistent-field (HF-SCF) and the 
valence correlation (evaluated at the frozen-core (fc) CCSD(T) level of theory) terms, 
using the formulas of \citet{Halkier-CPL99-HF} for the former and the two-parameter 
correction of \citet{Helga-JCP97-Basis} for the latter. 
At HF-SCF level of theory, the correlation consistent polarised basis sets 
cc-pV$n$Z ($n=\mrm{T},\mrm{Q},5$) of \citet{Dunn-JCP89-Gauss} and \citet{Woon-JCP95-CV} 
were used, while the fc-CCSD(T) calculations were carried out by using the cc-pVTZ and 
cc-pVQZ basis sets. 
For computing the contribution related to the core-valence (CV) electron correlation the 
difference between the all-electron (ae) and frozen-core results using the cc-pCVTZ basis 
set \citep{Woon-JCP95-CV} was used, while the contribution due to the diffuse functions was 
calculated by employing the aug-cc-pVTZ basis set\citep{Kendall-JCP92-basis}. 
All these terms were evaluated at CCSD(T) level of theory. 
The energies of these two structures were computed with the same approach but using the 
cc-pV$n$Z ($n=\mrm{Q},5,6$) basis sets. 
The vibrational corrections to both the equilibrium rotational constants were calculated 
at fc-MP2 level of theory \citep{M&P-PR34-manyel} and using the aug-cc-pVTZ basis set; 
the same level of theory provided also the anharmonic corrections used to compute the 
fundamental frequencies reported in Table~\ref{tab:excstates}.
The cubic force field data needed for determining the sextic centrifugal distortion constants 
were obtained at CCSD(T) level on the basis of its good accuracy reported in the literature 
\citep{PietroPC-JMS17-DFT}.  
Nuclear quadrupole coupling constants for the nitrogen atoms were computed at ae-CCSD(T) 
level of theory in conjunction with the pw-CV5Z basis set
\citep{Dunn-JCP89-Gauss,Peters-JCP02-basis} following the same procedure described previously
\citep{Cazzoli-JPC11-CHBrF2,PietroPC-JPC16-ClFEt}.

All the calculations carried out at CCSD(T) and MP2 level of theory were performed by using 
the CFOUR\footnote%
{
 CFOUR, Coupled-Cluster techniques for Computational Chemistry, a quantum-chemical program 
 package written by J.F. Stanton, J. Gauss, L. Cheng, M.E. Harding, D.A. Matthews, P.G. Szalay 
 et al., and the integral packages MOLECULE (J. Alml{\"o}f and P.R. Taylor), PROPS (P.R. Taylor), 
 ABACUS (T. Helgaker, H.J. Aa. Jensen, P. J{\o}rgensen, and J. Olsen), and ECP routines by 
 A. V. Mitin and C. van W{\"u}llen. For the current version, see \texttt{http://www.cfour.de}.
} 
suite of programs and its implementation of analytic second derivatives 
\citep{Gauss-CPL97-2ndDer}, while the sextic centrifugal distortion constants were computed 
using an appropriate suite of programs \citep{PietroPC-JMSt18-DFT} and the formulas reported 
in the literature \citep{Aliev-JMS76-CDC,Watson-JMS77-CDC,A&W-MR85-VibRot}.

A summary of the results provided by the present ab initio calculations is presented in 
Table~\ref{tab:geometry} together with the comparison with the values reported in the 
literature \citep{Sugie-JMS85-PGIM,Osman-IJMS14-PGIM}.

Concerning the equilibrium molecular parameters obtained by our calculations, on the basis 
of previous studies \citep[see, for example][]{Bak-JCP01-EqStr}, we may estimate an overall 
accuracy of 2--3$\times 10^{-3}$\,\AA\ for bond distances.
For comparison, the bond lengths reported by \citet{Sugie-JMS85-PGIM} listed in 
Table~\ref{tab:geometry} are generally shorter, while the ones computed by 
\citet{Osman-IJMS14-PGIM} (listed in the same Table) are generally longer. 
These discrepancies are due to the low level of theory used in these earlier studies.
As a matter of fact, the accuracy of molecular equilibrium structures is strongly 
dependent on the wave function method employed; for example, it is reported that 
computations carried out at HF level of theory generally underestimate the bond lengths,
while CCSD calculations, depending on the basis set used, may predict values that are either 
too short or too long \citep{Helga-JCP97-EqSr}.

As an independent test of the overall reliability of the theoretical calculations, one 
can apply the ab initio zero-point vibrational contributions to the experimentally 
derived ground state rotational constants and obtain the so-called ``semi-experimental'' 
equilibrium rotational constants. 
From these quantities the inertial defect can be estimated yielding 
$-6.7\times 10^{-4}$\,u\,\AA$^2$ for $Z$-PGIM and $-2.7\times 10^{-3}$\,u\,\AA$^2$ 
for $E$-PGIM.
These values are very close to zero as expected for a molecule with planar equilibrium 
configuration, thus indicating that both the electronic structures and the vibrational 
dynamics have been correctly modelled \citep{McCarthy-JCP16-HOCO}.

\begin{table*}[h]
 \caption[]{Theoretically computed  equilibrium geometries$^a$, dipole moments$^b$, 
            relative energy$^c$, and equilibrium rotational constants$^c$ of PGIM 
            isomers and comparison with previous works.
            \label{tab:geometry}}
 \footnotesize
 \begin{center}
 \begin{tabular}{ll D{.}{.}{4}D{.}{.}{4}D{.}{.}{4} c D{.}{.}{4}D{.}{.}{4}D{.}{.}{4}}
  \hline  \\[-1ex]
   & & \mcl{3}{c}{$Z$-PGIM}  &  & \mcl{3}{c}{$E$-PGIM} \\[0.5ex]
   \cline{3-5}\cline{7-9} \\[-1ex]
   & & \mcl{1}{c}{this work} & \mcl{1}{c}{Ref. (1)$^d$} & \mcl{1}{c}{Ref. (2)$^e$} & & 
       \mcl{1}{c}{this work} & \mcl{1}{c}{Ref. (1)$^d$} & \mcl{1}{c}{Ref. (2)$^e$} \\[1ex]
  \hline  \\[-1ex]
  \ce{C#C}  & /\AA               &     1.2074 &      1.1845 &   1.226 & &   1.2066   &     1.1838  &    1.225 \\
  \ce{C-C}  & /\AA               &     1.4374 &      1.4480 &   1.460 & &   1.4345   &     1.4444  &    1.457 \\
  \ce{C=N}  & /\AA               &     1.2767 &      1.2504 &   1.288 & &   1.2770   &     1.2506  &    1.289 \\
  \ce{H-C}  & /\AA               &     1.0635 &      1.0558 &   1.076 & &   1.0634   &     1.0557  &    1.076 \\
  \ce{C-H}  & /\AA               &     1.0863 &      1.0780 &   1.100 & &   1.0903   &     1.0827  &    1.102 \\
  \ce{N-H}  & /\AA               &     1.0199 &      1.0060 &   1.029 & &   1.0184   &     1.0046  &    1.028 \\[0.5ex]
  $\angle$(\ce{C-C-H}) & / deg   &   116.52   &    115.82   &         & &  115.54    &   114.60    &          \\
  $\angle$(\ce{H-C=N}) & / deg   &   118.11   &    118.38   &         & &  123.32    &   123.98    &          \\
  $\angle$(\ce{H-N=C}) & / deg   &   110.49   &    111.97   & 109.75  & &  109.92    &   110.94    &  109.06  \\      
  $\angle$(\ce{C-C=N}) & / deg   &   125.37   &    125.80   & 126.68  & &  121.14    &   121.42    &  121.00  \\[1ex]

  $\mu_a$   & / D                &     2.1449 &      2.39   &         & &    0.2567  &     0.23    &          \\
  $\mu_b$   & / D                &     0.1674 &      0.26   &         & &    1.9346  &     2.13    &          \\[1ex]
  $E_0$     & / kcal\,mol$^{-1}$ &     0.0    &     0.0     &   0.0   & &    0.8510  &     1.037   &    0.697 \\[1ex]
  $A_e$     & / MHz              & 54525.6766 & 56884.0     &         & &  63330.757 & 65775.0     &          \\
  $B_e$     & / MHz              &  4876.6255 &  4924.0     &         & &   4772.651 &  4833.0     &          \\
  $C_e$     & / MHz              &  4476.2794 &  4531.0     &         & &   4438.187 &  4502.0     &          \\[1ex]
  \hline\hline \\[-2ex]
 \end{tabular}
 \end{center}
 $^a$ Extrapolated ``best value'', CCSD(T)/CBS+CV.     \\[0.5ex]
 $^b$ Extrapolated, CCSD(T)/CBS+CV (absolute values).  \\[0.5ex]
 $^c$ See Appendix~\ref{sec:theor-details} for explanation.       \\[0.5ex]
 $^d$ Ref.~(1) is \citet{Sugie-JMS85-PGIM}, computed at HF/4-31G$^\ast$ level of theory. \\[0.5ex]
 $^e$ Ref.~(2) is \citet{Osman-IJMS14-PGIM}, computed at CCSD/aug-cc-pVDZ level of theory\@.
\end{table*}

\begin{table}[h]
 \caption{Anharmonic frequencies (cm$^{-1}$) for the singly-excited vibrational levels 
          of PGIM isomers$^a$.
          \label{tab:excstates}}
 \footnotesize
 \begin{center}
 \begin{tabular}{l c rr}
  \hline \\[-1ex]
   level & symmetry & \mcl{1}{c}{$Z$-PGIM} & \mcl{1}{c}{$E$-PGIM} \\[1ex]
  \hline \\[-1ex]
  $\nu_1$     & A$^\prime$ &  3322 &  3337 \\
  $\nu_2$     & A$^\prime$ &  3275 &  3286 \\
  $\nu_3$     & A$^\prime$ &  3028 &  2911 \\
  $\nu_4$     & A$^\prime$ &  2113 &  2121 \\
  $\nu_5$     & A$^\prime$ &  1592 &  1605 \\
  $\nu_6$     & A$^\prime$ &  1395 &  1385 \\
  $\nu_7$     & A$^\prime$ &  1224 &  1224 \\
  $\nu_8$     & A$^\prime$ &   922 &   916 \\
  $\nu_9$     & A$^\prime$ &   640 &   643 \\
  $\nu_{10}$  & A$^\prime$ &   591 &   596 \\
  $\nu_{11}$  & A$^\prime$ &   211 &   219 \\
  $\nu_{12}$  & A$^{\prime\prime}$ &  1099 &  1074 \\
  $\nu_{13}$  & A$^{\prime\prime}$ &   817 &   795 \\
  $\nu_{14}$  & A$^{\prime\prime}$ &   646 &   651 \\
  $\nu_{15}$  & A$^{\prime\prime}$ &   289 &   285 \\
  \hline\hline \\[-2ex]
 \end{tabular}
 \end{center}
 $^a$ Harmonic force field data corrected by cubic and quartic semi-diagonal 
      force constants obtained by fc-MP2 calculations (see text for details).
\end{table}

\bibliographystyle{aa}

\bibliography{jsc-astro,%
               astroch,%
               molphys,%
               cchains,%
               asymm,%
               instrument,%
               poly-ions,%
               misc}

\end{document}